%% file: main.tex
\documentclass[a4paper,11pt]{article}
\usepackage{jheppub} 
\usepackage{lineno}
\usepackage{graphicx}
\usepackage{subcaption}
\usepackage{framed}
\usepackage{amsthm}
\usepackage{hyperref}
\usepackage{xcolor}
\usepackage{amsmath}

\input{latexdefs_all}

\theoremstyle{definition}
\newtheorem{exmp}{Example}[section]

\makeatletter
\gdef\@fpheader{\vspace{0cm}}
\makeatother

\title{\boldmath Resurgence of large order relations}

\author[a,b]{Coenraad Marinissen}
\author[a]{, Alexander van Spaendonck}
\author[a]{ and Marcel Vonk}
\affiliation[a]{Institute of Physics, University of Amsterdam,
Science Park 904, 1090 GL Amsterdam, The Netherlands}
\affiliation[b]{Nikhef, Theory Group, Science Park 105, 1098 XG, Amsterdam, The Netherlands}

\emailAdd{c.b.marinissen@uva.nl}
\emailAdd{a.b.n.vanspaendonck@uva.nl}
\emailAdd{m.l.vonk@uva.nl}

\abstract{One of the main applications of resurgence in physics is the decoding of nonperturbative effects through large order relations. These relations connect perturbative asymptotic expansions of observables to expansions around other saddle points. Together, this data is unified in transseries that describe the nonperturbative structure. It is known that large order relations themselves also take the form of transseries. We study these large order transseries, uncover an interesting underlying geometry that we call the `Borel cylinder', and show that large order transseries in turn are resurgent -- that is: their nonperturbative sectors `know about each other' through Borel residues that are essentially equal to those of the original transseries. We show that with an appropriate resummation prescription, large order relations are often exact: they can be used to exactly compute perturbative coefficients -- not just their large order growth. Finally, we argue that Stokes phenomenon plays an important role for large order relations, for example if we want to extend the discrete index of the perturbative coefficients to arbitrary complex values.}

\begin{document}
\maketitle
\flushbottom

\section{Introduction}
\label{sec:intro}
One of the most common approximation methods for physics problems is perturbation theory. When a problem is too difficult to solve analytically, one introduces a small parameter $x$ or a large parameter $z$, and writes the answer to the problem in the form of a power series -- in this paper, usually of the form
\be\label{eq:perturbativeSeries}
 \varphi^{(0)}(z) = \sum_{g=0}^\infty \varphi^{(0)}_g z^{-g}\,.
\ee
Unfortunately, even as a method of approximation, this does not always solve the problem at hand: in many situations, the power series that one finds is not a convergent Taylor series for an analytic function, but it is {\em asymptotic} in the sense of Poincaré. That is, while every finite sum of its terms may give a better approximation of some function as $z$ gets larger, the coefficients $\varphi^{(0)}_g$ often grow factorially, $\varphi^{(0)}_g \sim A^{-g} \, g!$, and as a result the infinite sum converges for {\em no} value of $z$.

Nonetheless, a finite value can be assigned to these asymptotic series. The simplest method is that of optimal truncation where, for a given finite value of $z$, one sums all the terms in the series up to an `optimal' term $\varphi^{(0)}_G z^{-G}$ determined by $G \simeq |A z|$. However, this method is in general only accurate up to exponentially small contributions of the form $\sim \re^{-|Az|}$ which are {\em nonperturbative} in the expansion parameter $1/z$. In order to {\em resum} asymptotic series beyond the resolution of order $\re^{-|Az|}$, we can turn to the method of Borel summation \cite{Borel:1899}. This summation technique plays a central role in the theory of resurgence introduced by J. Écalle in the 1980s \cite{ecalle1985fonctions}, and contrary to optimal truncation provides a way to resum asymptotic series up to arbitrary precision. In fact, Écalle showed that Borel summation lies at the basis of a systematic procedure for finding missing nonperturbative contributions, and that these are intricately connected to one another in an algebraic sense that he formulated in his \textit{alien calculus}.

Over the past decade or so, techniques from resurgence have become more popular among physicists and have been applied to various physical models, including (but definitely not limited to) the Painlevé I equation \cite{Garoufalidis:2010ya, Aniceto:2011nu, Baldino:2022aqm, vanSpaendonck:2022kit}, quantum mechanics \cite{AIHPA_1983__39_3_211_0, AIHPA:1999:71, Zinn-Justin:2004vcw, Dunne:2014bca, vanSpaendonck:2023znn}, quantum field theories \cite{Dunne:2012ae, Gukov:2016njj,  Marino:2019eym, DiPietro:2021yxb}, hydrodynamics \cite{Heller:2015dha, Aniceto:2015mto}, two-dimensional gravity \cite{Gregori:2021tvs, Eynard:2023qdr}, topological strings \cite{marino2008nonperturbative, Marino:2008ya, Couso-Santamaria:2013kmu, Gu:2023mgf} and many more. In many of these examples, resurgence analysis suggests that we should supplement asymptotic series like \eqref{eq:perturbativeSeries} with additional nonperturbative contributions, which can be united in what is known as a {\em transseries} (see e.g.~\cite{edgar2010transseries}), in its simplest form:
\be
\Phi(\vec{\gs}, z) 
= \sum_{n=0}^\infty \gs_n \, \varphi^{(n)}(z) \, \re^{-nAz}\,.
\label{eq:firsttransseries}
\ee
The formal power series $\varphi^{(n)}(z) = \sum_{g=0}^\infty \varphi^{(n)}_g z^{-g}$ are often called \textit{transseries sectors} and they are usually all asymptotically divergent -- including our original series \eqref{eq:perturbativeSeries}, of course. From a physics point of view, one can often think of the transseries as the semiclassical expansion of a path integral which yields a sum over saddle point contributions. In this interpretation, $z$ is the inverse of the reduced Planck constant $\hbar$ and $nA$ is the instanton action of the $n$-th saddle point solution. In the transseries, each saddle point contribution is weighted by a parameter $\gs_n$ whose value is determined by boundary conditions imposed on the path integral.

A well-known result from resurgence is that knowledge of the perturbative coefficients $\varphi^{(0)}_g$ in \eqref{eq:perturbativeSeries} allows us to decode the coefficients in other sectors $\varphi^{(n)}$ of the transseries. In practice, this is achieved through the use of so-called \textit{large order relations}, which in the literature often appear in the form
\begin{equation}\label{eq:LORintro}
\frac{\varphi_g^{(0)}}{\Gamma(g)}
\simeq \sum_{k=1}^\infty \frac{S_1^k}{2\pi\ri}  
\sum_{h=0}^\infty(kA)^{g-h} \frac{\Gamma(g-h) }{\Gamma(g)} \varphi_h^{(k)}\,,
\end{equation}
and which should be understood to hold in the large $g$ limit. This form of the large order relation -- in which each `sector' $k$ carries a constant of proporionality $S_1^k$, with $S_1$ a Stokes constant -- was presented first in \cite{Aniceto:2011nu} in the contexts of ODEs and matrix models, and is a multi-instanton generalization of the more rigorously derived large order relations found in \cite{Costin_1999, garoufalidis2009universality}.

The study of the large order behaviour of asymptotic series in physics predates the conception of resurgence and large order relations. In the seventies, Bender and Wu studied the large $n$ behaviour of the perturbative energy levels $E_n$ of the quartic anharmonic oscillator~\cite{Bender:1969si} and showed that their large order behaviour could be recovered from studying the one-instanton sector of the `negatively coupled' quartic oscilator~\cite{Bender:1973rz}. Many other implicit and explicit discussions of large order behaviour have appeared over the years, e.g.\ \cite{PhysRevD.16.408, SHikami_1979, LeGuillou:1990nq}. To the best of our knowledge, the first (leading instanton) large order formulas of the form \eqref{eq:LORintro} appeared in \cite{Costin_1999} and subsequently also in \cite{marino2008nonperturbative, Marino:2008ya, garoufalidis2009universality} before being generalized to the multi-instanton version \eqref{eq:LORintro} from~\cite{Aniceto:2011nu}. 

In the more recent mathematics literature, the focus of resurgence research has not been on large order behaviour as much as it has been in the physics literature, but Écalle and Sharma described a method to study large order relations in \cite{zbMATH05976778}, and most of the essectial techniques needed to discuss large order behaviour can be found in the textbook \cite{sauzinbook}. In 2020, Sauzin used these techniques to study the large order behavior of the asymptotic `Stirling series' that occurs when expanding the logarithm of the gamma function \cite{sauzinprivate}.

\begin{figure}
\centering
\subfloat[]{
    \includegraphics[width=.47\textwidth]{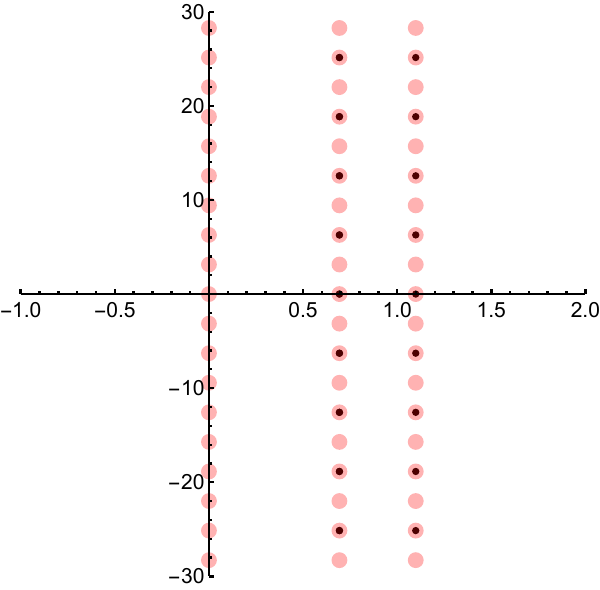}
    \label{fig:BPplotAdler1}
}
\subfloat[]{
    \includegraphics[width=.47\textwidth]{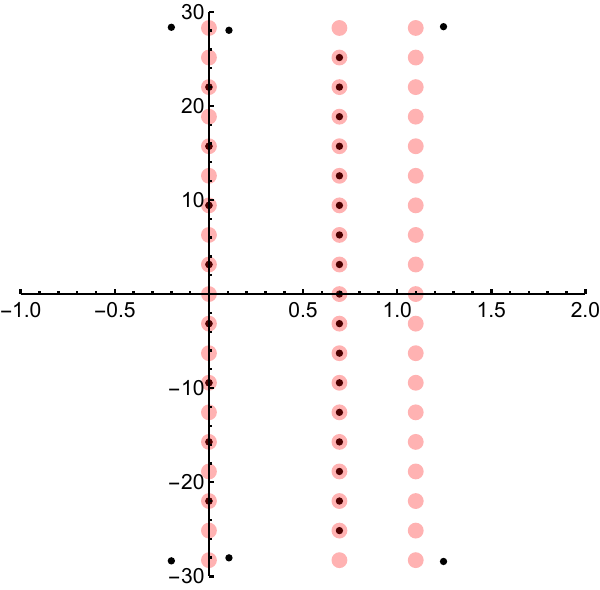}
    \label{fig:BPplotAdler2}
}
\caption{Borel transform of the Adler function large order expansion. The black dots are the singularities of the Borel-Padé transform of the $1/g$ expansion of~\eqref{eq:LORintro} for $k=1$, as discussed in~\cite{Laenen:2023hzu}. The red transparent disks are added to aid the eye and are positioned at $\log(\ell)+\pi\ri n$, with $\ell=1,2,3$ and $n\in \mathbb{Z}$. The two plots show the contributions to the large order relation coming from the two leading nonperturbative sectors. The spurious black dots with imaginary part $\sim \pm 9 \pi \ri$ in the right hand plot are a result of numerical instabilities.}
\label{fig:BPplotAdler}
\end{figure}

The applications of large order relations go beyond toy models: in a recent investigation of the resurgent structure of the Adler function of QED and QCD \cite{Laenen:2023hzu}, large order relations were studied with the aim of extracting higher nonperturbative sectors that have an interpretation as renormalon contributions. For each $k$ in \eqref{eq:LORintro} the ratio of gamma functions yields a new asymptotic series in $1/g$, which in the context of the Adler function (as in many other examples of asymptotic series) was studied by plotting poles of the Padé approximant to its Borel transform. This led to the plots shown in  figure~\ref{fig:BPplotAdler}, where the black dots indicate the singularities of the Borel transform of these series, which in turn should determine nonperturbative contibutions to the large order relation. It was the interesting singularity structure observed there -- somewhat reminiscent of the `peacock patterns' observed in \cite{Garoufalidis:2020xec} -- that led us to the question which underlying structure gives rise to these patterns.

\begin{figure}
\centering
\subfloat[]{
    \includegraphics[width=.46\textwidth]{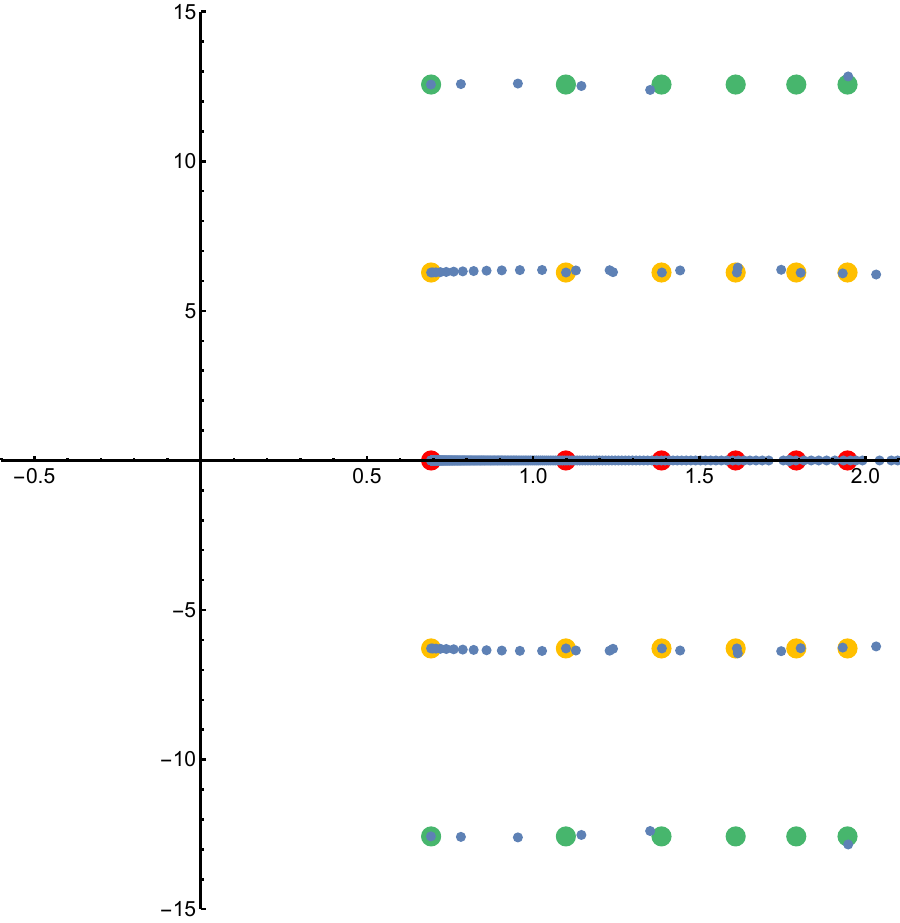}
    \label{fig:BP_plot_free_energy}
}
\subfloat[]{
    \includegraphics[width=.48\textwidth]{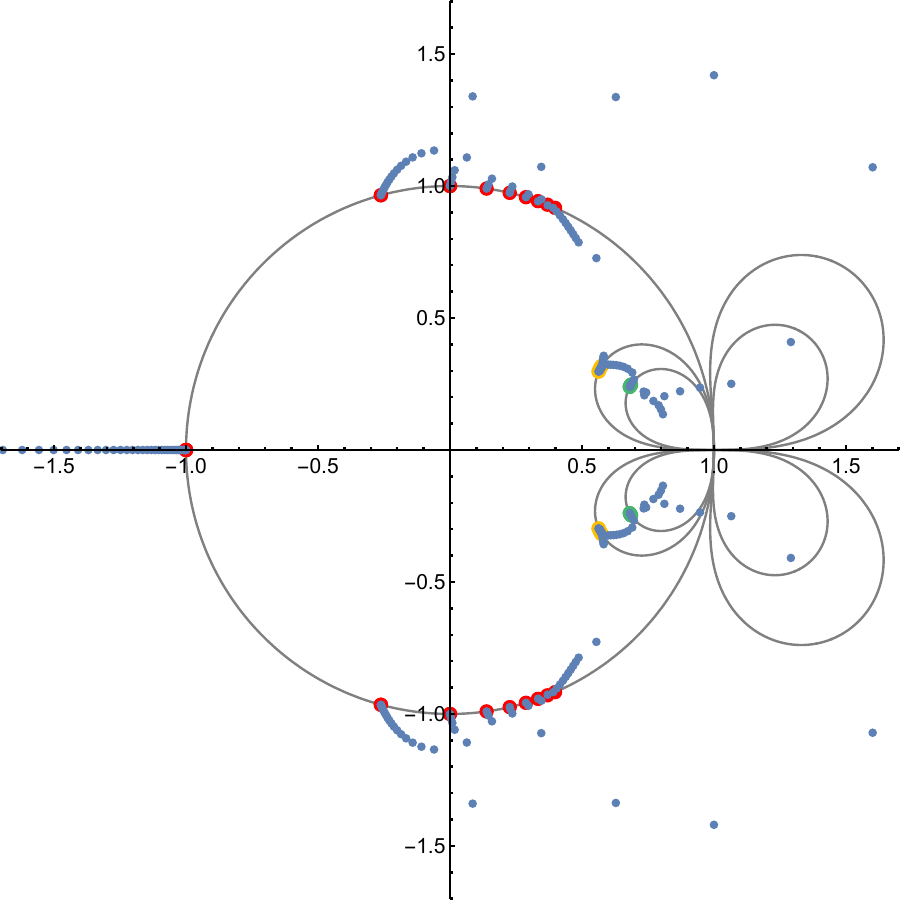}
    \label{fig:CBP_plot_free_energy}
}
\caption{Figure~\ref{fig:BP_plot_free_energy} shows the singularities of the Borel-Padé transform of~\eqref{eq:LORintro} for $k=1$ for the quartic free energy. Similar to the plots shown in figure~\ref{fig:BPplotAdler}, we observe singularities at $\log(\ell)+2\pi\ri n$, but now for integers $\ell\geq 2$ and $n\in\mathbb{Z}$. As the singularities are branch cuts (mimicked by the accumulation of poles of the Padé approximant), the poles corresponding to different branch cuts are hard to distinguish. In figure~\ref{fig:CBP_plot_free_energy}, we therefore show the Padé singularities after applying a conformal map to the Borel plane~\cite{Costin:2020hwg,Costin:2021bay}, which maps the singularities on the positive real axis to the unit circle; the images of some other lines with imaginary part $2 \pi n$ are also drawn. We clearly observe that the branch cuts are pulled apart and can therefore conclude that the poles on the real axis in figure~\ref{fig:BP_plot_free_energy} belong to different branch cuts. We refer to example~
\ref{ex:AdlerCylinder} for more details.}
\label{fig:BPplotFreeEnergy}
\end{figure}

The observed pattern is not specific to the Adler function: it also appears for a variety of other models -- see e.g. figure~\ref{fig:BPplotFreeEnergy}, which displays a similar plot for a toy model that we will study extensively later in this paper. In this case, the Padé poles line up to mimic branch cuts (rather than poles) in the Borel transform of the large order expansion, but the repeating structure of singularities is essentially the same. These observations are particularly interesting in light of the fact, as can be seen from \eqref{eq:LORintro}, that the $1/g$ expansion of the large order relation~\eqref{eq:LORintro} {\em itself} takes the form of a transseries \cite{Aniceto:2011nu, Aniceto_2019}. These two ingredients form the starting point of this work.

In this work we study the resurgent structure of the transseries that large order relations constitute. 
We show how this resurgent structure can be understood using the so-called Stirling transform, which repackages all the coefficients of the instanton sectors in \eqref{eq:firsttransseries} into new series that appear in the large order relation \eqref{eq:LORintro}. In particular, these new series make the repeated pattern in the imaginary direction in figures \ref{fig:BPplotAdler} and \ref{fig:BP_plot_free_energy} manifest, leading to an underlying geometric structure that we dub the {\em Borel cylinder}. We subsequently derive the Stokes phenomenon \cite{stokes_2009a, stokes_2009b} of the large order transseries, explain how it relates to the Stokes phenomenon of the original asymptotic expansion, and show how one can consistently resum the large order transseries. In doing so, we test to what extent the resummation of the large order relation describes the perturbative coefficients {\em exactly}: when we resum ~\eqref{eq:LORintro} for finite $g$, can we calculate the perturbative coefficients $\varphi^{(0)}_g$ to arbitrary precision? We will find that the answer to this question is often `yes', even in cases where one ignores potential contributions to the large order relation coming from a Cauchy integral at infinity. Finally, we will see that the resummation establishes a natural extension of the large order relation to non-integer and even complex values of $g$. Throughout the paper, we provide numerous examples to illustrate various concepts and to numerically verify our statements.

In section \ref{sec:backrgound}, we review the tools from resurgence that are necessary for the study of large order relations. The reader who is already familiar with the standard ingredients of resurgence -- in particular Borel summation, transseries and the usual derivation of large order formulas -- may safely skip this section, up to perhaps a quick look at the subtleties in the definition of Borel residues that we point out in section \ref{sec:AmbStokesData}. In section \ref{sec:resurgenceofLOR}, we study the large order relations and their transseries structure. In section \ref{sec:StirlingTransform} we introduce the Stirling transform and its Borel transform, which in section \ref{sec:ResurgenceStructureLOR} allow us to establish resurgence relations for the large order transseries. In section \ref{sec:StokesAutoLOR}, we then derive the Stokes automorphism and a consistent way for resumming these transseries, and finally formulate `exact' large order relations in section \ref{sec:ELOR}. Section \ref{sec:ELORforPartitionFunction} is devoted to a test of the exact large order relation for a toy model called the quartic partition function and section \ref{sec:conclusion} finishes with our conclusions and an outlook. There are three appendices: in appendix~\ref{sec:nBorel} we provide some more background on the various possible Borel transforms and Borel summations, in appendix~\ref{sec:AppBorelStirling} we discuss the Stirling transform, and in appendix~\ref{sec:AppGamma} we give a derivation of the Stokes phenomenon for the gamma function that we need in section \ref{sec:LORforPartitionFunction}.

\section{Resurgence and asymptotics}
\label{sec:backrgound}
In this section, we review material from the theory of Borel summation, transseries and resurgence that we need in the rest of the paper. In particular, we collect the ingredients that usually go into the derivation of a large order formula. We mostly use the notation of the primer \cite{Aniceto_2019}, to which we also refer for further background information. Another useful reference in which the material of sections \ref{sec:borel} -- \ref{sec:AmbStokesData} is explained in a more mathematical language is \cite{sauzinbook}.

In section \ref{sec:borel}, we review the notion of (Gevrey-1) asymptotic series, and how these divergent expansions can be Borel summed into honest functions. In section \ref{sec:stokes} we recall how ambiguities in the Borel summation procedure, originating from Stokes lines, lead to the concepts of Stokes phenomenon and transseries. We point out an important but often overlooked subtlety in the description of Stokes phenomenon -- more precisely, in the definition of the so-called Borel residues -- in section \ref{sec:AmbStokesData}. Finally, in section \ref{sec:classicallargeorder}, we use the concepts that we have introduced to review the `classical' derivation of the large order formula for a transseries.

\subsection{Borel summation of asymptotic expansions}
\label{sec:borel}
Assume that some physical problem provides us with a power series
\be
 \varphi^{(0)}(z) = \sum_{g=0}^\infty \varphi^{(0)}_g \, z^{-g},
 \label{eq:asymptotic}
\ee
where, as is common in the resurgence literature, we have assumed an expansion in nonnegative powers around large $z$.  This assumption is not very restrictive: by changing variables and multiplying by an appropriate power of $z$ we can bring any power series in this form. We will further assume that this series is asymptotic, of the {\em Gevrey-1} type -- that is: we assume that as $g \to \infty$, the coefficients $\varphi^{(0)}_g$ grow factorially as
\be
 \frac{\varphi^{(0)}_g A^{g+\beta}}{\Gam(g+\beta)} \sim c + \cO(1/g),
 \label{eq:gevrey1}
\ee
for some nonzero complex constants $A, c$ and a real constant $\beta$. In most of what follows, we will assume for simplicity that $\beta$ is an integer. The results in this paper are easily generalized to the case of noninteger $\beta$.

Due to the factorial growth of the coefficients $\varphi_g^{(0)}$ in (\ref{eq:gevrey1}), the series (\ref{eq:asymptotic}) does not converge for any nonzero value of $z^{-1}$. When (\ref{eq:asymptotic}) arises in a physics setting this is a problem, as one would expect the series $\varphi^{(0)}(z)$ to not just be purely formal but to describe a {\em function} that has a finite value for some range of $z$-values. The usual way to cure this problem is to take the {\em Borel transform} of the series:
\be
 \cB[\varphi^{(0)}](t) \equiv \sum_{g=0}^\infty \frac{\varphi^{(0)}_g}{g!} \, t^g.
 \label{eq:boreldef}
\ee
Dividing the coefficients of our original series by $g!$ greatly improves the convergence of the series: the new series now has a radius of convergence equal to $|A|$. In many cases, $\cB[\varphi^{(0)}](t)$ can in fact be analytically continued to the whole complex $t$-plane (or a multiple cover thereof -- something we will come back to in more detail soon) minus a discrete set of points where the analytic continuation of $\cB[\varphi^{(0)}](t)$ has singularities. We will assume that we are in such a situation, and will write $\cB[\varphi^{(0)}](t)$ for the function obtained from this analytic continuation as well.

\begin{exmp}
\label{ex:borellog}
\begin{leftbar}
Let us consider the case where $\varphi^{(0)}_0 = 0$ and $\varphi^{(0)}_g = - A^{-g} (g-1)!$ for $g>0$. This is clearly of the form (\ref{eq:gevrey1}) with $\beta=0$, $c=-1$, and all $\cO(1/g)$ corrections vanishing in this example. The Borel transform of $\varphi^{(0)}(z)$ is then
\be
 \cB[\varphi^{(0)}](t) = - \sum_{g=1}^\infty \frac{1}{g}\left(\frac{t}{A}\right)^g = \log\left(1-\frac{t}{A}\right),
\ee
where the series after the first equals sign converges for $|t|<|A|$, and the second equals sign contains the analytic continuation of this convergent series into a multi-valued function on the entire $t$-plane except for $t=A$, where the function has a logarithmic branch point. Equivalently, we can replace the $t$-plane by its $\bZ$-multiple cover branched at $t=A$, which is the Riemann surface on which the logarithm is single-valued. 
\end{leftbar}
\end{exmp}

When the Borel transform (\ref{eq:boreldef}) can indeed be analytically continued to most of the complex plane -- more precisely: along paths $\gam$ that start at the origin and go off to infinity -- the transform has a well-known `inverse': one can define a function
\be
 \cS_\gam\varphi^{(0)}(z) \equiv z \int_\gam \cB[\varphi^{(0)}](t) \, e^{-tz} \, dt
 \label{eq:laplacedef}
\ee
known as a {\em Borel sum} (along a path $\gam$ for which the integral converges) of the formal power series $\varphi^{(0)}$. Usually, we will take $\gam$ to be a straight, semi-infinite ray that has an angle $\theta$ with the positive real axis; in such a case we denote the Borel sum by $\cS_\theta\varphi^{(0)}$. Of course, when $\cB[\varphi^{(0)}]$ has a singularity along the ray in the direction $\theta$, this Borel sum is not well-defined; in these cases (whenever it is clear what the singular direction $\theta$ is) we denote the {\em lateral Borel sums} with angles infinitesimally larger or smaller than $\theta$ by $\cS_+\varphi^{(0)}$ and $\cS_-\varphi^{(0)}$ respectively.

The Laplace transformation introduced above is not literally the inverse of the Borel transform: while $\varphi^{(0)}$ is a formal (and divergent) power series, $\cS_\gam\varphi^{(0)}$ is an honest function -- so the two are clearly not `the same object'. However, if one were to plug (\ref{eq:boreldef}) into (\ref{eq:laplacedef}), exchange the sum and integral (an operation which is not usually allowed!), and perform the integrals -- say with $\gam$ the positive real axis -- then one {\em would} find back the original asymptotic series. It is in this sense that the Laplace transform (\ref{eq:laplacedef}) is the `inverse' of the Borel transform (\ref{eq:boreldef}).

In fact, the Borel transform followed by the Laplace transform solves exactly the problem we mentioned: it turns an asymptotically divergent series in $z$ into an honest function of $z$, and one can show that for Gevrey-1 series a particular asymptotic expansion of that function at large $z$ is the original expansion $\varphi^{(0)}(z)$ that we started with. While this solves our problem, this is certainly not the end of the story: the construction depends on the integration path $\gam$ in the Laplace transform, and therefore the function one arrives at is generally not unique. As it turns out, this non-uniqueness is a virtue rather than a problem.

\subsection{Stokes phenomenon}
\label{sec:stokes}
The existence of many different inverse Borel transforms is the starting point of the theory of resurgence. Note that if two paths $\gam$ can be deformed into one another without crossing any singularities of $\cB[\varphi^{(0)}](t)$, then by Cauchy's theorem they lead to the same Borel sum. (This assumes for now the absence of contributions at infinity -- we will discuss the more general situation extensively in sections \ref{sec:ELOR} and \ref{sec:ELORdiscussion}.) As a result, the Borel sums $\cS_\theta\varphi^{(0)}$ do not change when $\theta$ is varied until $\theta$ crosses a ray on which $\cB[\varphi^{(0)}](t)$ has singularities. These special rays in the Borel plane are known as {\em Stokes lines}\footnote{In the literature, Stokes lines are often defined to lie in the $z$-plane, rather than the Borel $t$-plane. This is indeed equivalent when we \textit{fix} the contour along we resum, as we explain in appendix \ref{sec:nBorel}. }.

In most of what follows, we will be interested only in what happens at a single Stokes line. Without loss of generality, by rotating the complex $z$- and $t$-variables if necessary, we may assume that this Stokes line is the positive real axis. The Stokes line can contain more than a single singularity, but we do make the assumption that any singularities on it are at $t = \ell A$ for $\ell \in \bZ_{>0}$. This is often a reasonable assumption; for example, there may be multi-instanton configurations in our physical problem, and those would lead to precisely such a pattern of singularities.

We now further assume that we are in the setting of {\em simple resurgent functions}, where the Borel transform only has poles and logarithmic branch cuts as its singularities. We will in particular be interested in the case of logarithmic singularities, but this is not a very strong restriction, as the following variation on example \ref{ex:borellog} shows:

\begin{exmp}
\label{ex:borelpole}
\begin{leftbar}
Let us consider the case where $\varphi^{(0)}_0 = 0$ and $\varphi^{(0)}_g = - A^{-g} g!$ for $g>0$. The only difference with example \ref{ex:borellog} is that we replaced $(g-1)!$ by $g!$. The coefficients are again of the form (\ref{eq:gevrey1}), now with $\beta=1$ and $c = A$. The Borel transform of $\varphi^{(0)}(z)$ is now
\be
 \cB[\varphi^{(0)}](t) = - \sum_{g=1}^\infty \left(\frac{t}{A}\right)^g = -\frac{t}{A-t},
\ee
where again the second equals sign contains the analytic continuation. 
\end{leftbar}
\end{exmp}

Comparing examples \ref{ex:borellog} and \ref{ex:borelpole}, we see that the difference between a pole and a logarithmic branch cut in the Borel plane lies in an asymptotic coefficient growth proportional to $(g-1)!$ and $g!$ respectively -- or equivalently, in a different value of $\beta$ in (\ref{eq:asymptotic}). One can turn this observation around: if we replace the standard Borel transform (\ref{eq:boreldef}) by the `$n$-Borel transform' defined as
\be
 \cB_n[\varphi^{(0)}](t) \equiv \sum_{g=0}^\infty \frac{\varphi^{(0)}_{g+n}}{g!} \, t^{g},
 \label{eq:nboreldef}
\ee
then by taking $n$ sufficiently small (and declaring $\varphi^{(0)}_h=0$ for $h<0$ if needed) one can make sure that at a given point $t=\ell A$ in the Borel plane only a logarithmic branch cut appears, and no pole. Just like the ordinary Borel transform, the $n$-Borel transform has a Laplace transform that acts as its inverse (in the sense described above), and we can work with it in a completely analogous way to the ordinary transform. We explain this in some more detail in appendix \ref{sec:nBorel}. Because of these observations, in most of what follows we will assume that we are in a situation with only logarithmic branch cuts in the Borel plane, essentially without loss of generality. The Borel transform that we use in practice will usually be the Borel transform~$\cB_0$, and we will leave out the 0-index.

Let us then assume that near some singularity $t=\ell A$, the Borel transform of $\varphi^{(0)}(z)$ has the form
\be
 \cB[\varphi^{(0)}](t) = \frac{\fS_{0\to \ell}}{2 \pi i} \  \cB[\varphi^{(\ell)}](t-\ell A) \ \log(t-\ell A)+\text{regular},
 \label{eq:BorelResurgence}
\ee
where `regular' stands for a function that is analytic (and therefore non-singular) at $t=\ell A$. Here, we have written the prefactor of $\log(t-\ell A)$ in a particular form suggested by resurgence: after extracting a constant factor $\frac{\mathsf{S}_{0\to \ell}}{2\pi i}$, we write this prefactor as the Borel transform of a new asymptotic series $\varphi^{(\ell)}$ whose meaning and importance will become clear momentarily. Of course, there is some freedom in scaling $\mathsf{S}_{0\to \ell}$ and $\varphi^{(\ell)}$ in opposite ways; usually this freedom is fixed by, for example, requiring that the leading term in $\varphi^{(\ell)}$ is unity. 

\begin{figure}
    \centering
    \includegraphics[width = 0.9 \linewidth]{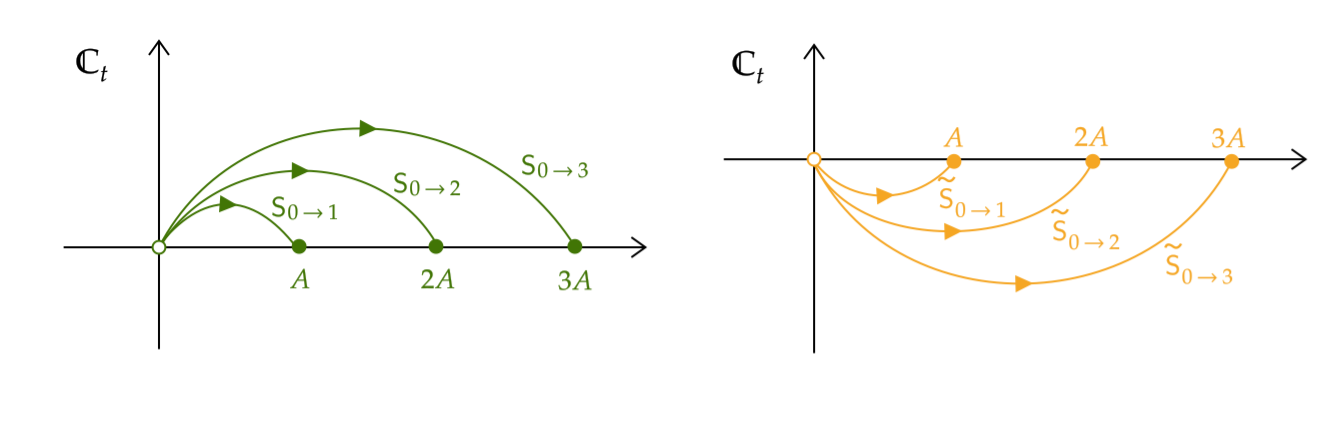}
    \caption{Conventions for the description (\ref{eq:BorelResurgence}) of the Borel transforms near singularities at $t=\ell A$: on the left the standard convention (continuation above the Stokes line) and on the right an alternative convention (continuation below the Stokes line). }
    \label{fig:residueconvention}
\end{figure}

The constants $\mathsf{S}_{0\to \ell}$ are called {\em Borel residues}. An important observation is that they are formally \textit{multivalued}, since there are in general many paths $\gam$ along which we can analytically continue the Borel transform from $t=0$ to $t = \ell A$. Therefore, to properly define the $\mathsf{S}_{0\to \ell}$, we must commit to a convention with regards to these paths. A standard convention is the one shown in figure \ref{fig:residueconvention} on the left side, where we we analytically continue the Borel transform above the Stokes line to the singularity of interest. We will come back to this ambiguity in section \ref{sec:AmbStokesData}.

\begin{figure}
    \centering
    \includegraphics[width= 1 \linewidth]{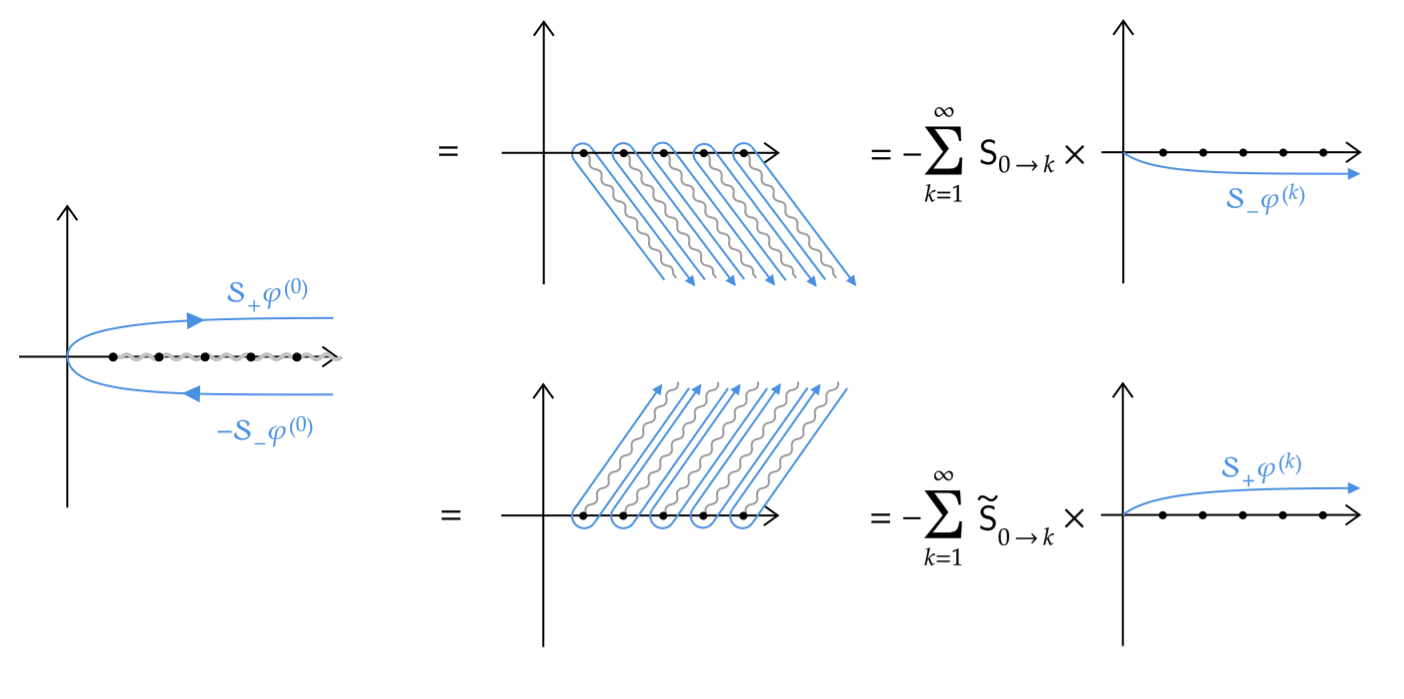}
    \caption{Considering the \textit{difference} of lateral resummations, there are  two natural choices: we either let all branch cuts (and thus the Hankel contours) run off to infinity below the real line (upper graphs) or we let them run off to infinity above real line (lower graphs).}
    \label{fig:TwoHankels}
\end{figure}

Whenever we encounter a Stokes line, we will be interested in computing the discontinuity
\be
    \text{disc} \, \varphi^{(0)}(z) = (\cS_+-\cS_-)\varphi^{(0)}(z).
\ee
In computing this discontinuity, it is important that the Laplace integrals for both $\cS_+$ and $\cS_-$ do not cross any branch cuts associated to the logarithmic contributions to the Borel transform (\ref{eq:BorelResurgence}). One may achieve this by aligning all branch cuts with the Stokes line (see the left hand side of figure \ref{fig:TwoHankels}), but a more conventient choice is to let all branch cuts go off to infinity in a diagonal direction and deform the integration contour into a sum of so-called {\em Hankel contours}, as depicted in the middle of figure \ref{fig:TwoHankels}. 

By decomposing the integration contour in terms of Hankel contours around individual singularities in this way, we can use the resurgence relations (\ref{eq:BorelResurgence}) to find\footnote{Here, we have exchanged the order of summation over $k$ and integration in $t$. This is an operation that, especially in a resurgence setting, has to be treated with care, but since the sums over $k$ (as opposed to those over $g$) generally have a nonzero radius of converge, this exchange of sums and integrals can often be put on solid footing.}
\be
(\cS_+-\cS_-)\varphi^{(0)}(z) = - \cS_- \left(\sum_{k=1}^\infty  \fS_{0\to k} \,\varphi^{(k)}(z)\, \re^{-kAz}\right). 
\label{eq:discontinuity}
\ee
More suggestively, we can write this as
\be
\cS_+\varphi^{(0)}(z) = \cS_- \left(\varphi^{(0)}(z) - \sum_{k=1}^\infty  \fS_{0\to k}\, \varphi^{(k)}(z) \,\re^{-kAz}\right), 
\ee
which tells us that the Borel summation along one side of the Stokes line is equal to the resummation along the other side when we modify the asymptotic expansion in the appropriate way. This is a manifestation of {\em Stokes phenomenon}. Using the {\em Stokes automorphism} $ \mathfrak{S}$ defined by
\be
    \cS_+ = \cS_- \circ \mathfrak{S},
    \label{eq:StokesAutDef}
\ee
we can write an equivalent statement in terms of asymptotic expansions which reads
\be
\mathfrak{S}\varphi^{(0)}(z) = \varphi^{(0)}(z) - \sum_{k=1}^\infty  \fS_{0\to k}\, \varphi^{(k)}(z)\, \re^{-kAz}.
\label{eq:StokesJump1}
\ee
An important result in the theory of resurgence is that the relation (\ref{eq:BorelResurgence}) generalizes to the other formal power series $\varphi^{(k)}$ as well: we have that
\be
 \cB[\varphi^{(k)}](t) = \frac{\fS_{k\to k+\ell}}{2 \pi i} \  \cB_0[\varphi^{(k+\ell)}](t-\ell A) \, \log(t-\ell A)+\text{regular},
 \label{eq:BorelResurgenceGeneral}
\ee
for integers $k,\ell$ such that $k\geq 0$ and $k+\ell \geq 0$. We will use the terminology that the sectors {\em know about each other}. (We prefer not to use the term `see each other', as the paths along which the Borel residues $\fS_{k\to k+\ell}$ are determined are generally not straight lines -- see figure \ref{fig:residueconvention}.) As a consequence, the action of the Stokes automorphism \eqref{eq:StokesJump1} generalizes to
\be
\mathfrak{S}\varphi^{(k)}(z) = \varphi^{(k)}(z) - \sum_{\ell=1}^\infty  \fS_{k\to k+\ell}\, \varphi^{(k+\ell)}(z)\, \re^{-\ell Az}\, ,
\label{eq:StokesJump2}
\ee
for generic sectors $\varphi^{(k)}$. The above expressions \eqref{eq:StokesJump1} and \eqref{eq:StokesJump2} show that it is useful to unite the formal, asymptotic power series $\varphi^{(n)}$ in a so-called {\em transseries},
\be
    \Phi(\vec{\gs}, z) = \sum_{n=0}^\infty \gs_n \, \varphi^{(n)}(z) \, \re^{-nAz},
    \label{eq:transseriesdef}
\ee
which is a double formal expansion, both in the perturbative expansion parameter $z^{-1}$ and in the nonperturbative {\em transmonomial} $\re^{-Az}$. In the above expression we have assigned weigths $\gs_n$ -- called {\em transseries parameters} -- to each {\em transseries sector} $\varphi^{(n)}$; these are all collected in the single vector $\vec{\gs}$ on the left hand side. The advantage of introducing these parameters is that we can now also view any asymptotic expansion before the Stokes jump \eqref{eq:StokesJump2} -- that is, simply $\varphi^{(k)}$ -- as a transseries with $\gs_n=\delta_{n,k}$. In fact, one may be more general and think of the Stokes automorphism as mapping one transseries expansion -- with particular values of $\gs_n$ and summed above the Stokes line using the $\cS_+$ procedure -- to another one -- with other values of $\gs_n$ and summed below the Stokes line using the $\cS_-$  procedure.

Using the transseries notation, we can write the action of the Stokes automorphism as an action on the transseries parameters:
\be
    \mathfrak{S} \Phi(\vec{\gs}, z) =  \Phi(\mathbb{S} \cdot \vec{\gs}, z)
    \label{eq:StokesJump3}\, .
\ee
Effectively, we are multiplying the (infinite) vector $\vec{\gs}$ with an (infinite) lower-triangular \textit{Stokes matrix} $\mathbb{S}$ whose elements are given by the Borel residues
\be
    \left(\mathbb{S}\right)^n_{\ m} = \begin{cases}
        - \fS_{m \to n} &  \text{for } m<n \\
        1 & \text{for } m=n \\
        0 & \text{for } m>n \, ,\\
    \end{cases} \qquad  = \quad \begin{pmatrix}
        1 & 0  & 0 &\ldots\\
        - \fS_{0\to1}\ & 1 & 0 &\ldots\\
        - \fS_{0\to2}\ & - \fS_{1\to2}\ & 1 &\ldots\\
        \ldots & \ldots & \ldots\  &  \ldots 
    \end{pmatrix}
    \label{eq:StokesMatrix}
\ee
for integers $m,n \geq 0$. In particular, this means that we have the following action on a single transseries parameter
\be
     \mS: \gs_{n} \mapsto \gs_{n} - \sum_{m=0}^{n-1} \gs_m \fS_{m \to n}\,.
    \label{eq:Stokesactionpar}
\ee
This generic description expresses the Stokes phenomenon in terms of an \textit{infinite} set of Borel residues. However, once we specify the underlying problem and compute the Borel residues, we often find that they depend on a \textit{finite} set of constants called \textit{Stokes constants}. Moreover, in such cases we generically find that the number of independent transseries parameters $\gs_n$ reduces to a finite number as well, as the following example shows. 
\begin{exmp}
\label{ex:StokesAutoFreeEnergy}
    \begin{leftbar}
        In section \ref{sec:ELORforPartitionFunction} we will introduce the \textit{quartic free energy} -- a popular toy model to study resurgence (see e.g. \cite{Aniceto_2019}). Throughout this section and the next, we will occasionally refer forward to aspects of this example to clarify the material. The quartic free energy has a two-parameter transseries solution that can be written as 
        \be
            F(x, \gs, \rho)=\sum_{n=0}^\infty \gs^n F^{(n)}(x) \re^{-nA/x}+\rho\, ,
        \ee
        where the $F^{(n)}(x)$ are formal power series in $x$. Their Borel transforms have a Stokes line along the positive real axis and therefore $F(x, \gs, \rho)$ undergoes a Stokes transition. Despite the fact that there are an infinite number of transseries sectors $F^{(n)}$, it turns out that setting $\gs_n = \gs^n$ is a convenient parametrization under which the Stokes automorphism along the $\arg(x) = 0$ direction acts as
        \be
        \mathfrak{S}_0 F(x, \gs, \rho) = F(x, \gs + S_1, \rho)\,.
        \label{eq:quarticFEex}
        \ee
        Thus we can reduce the number of degrees of freedom for this particular transition to \textit{one} parameter $\gs$ and \textit{one} Stokes constant $S_1$. These relate to our generic formulation \eqref{eq:transseriesdef} and \eqref{eq:StokesJump2} when
        \be
            \fS_{k\to k+l}= - \binom{k+l}{k} S_1^l\,.
            \label{eq:ResidueQFE1}
        \ee
        The second parameter $\rho$ is irrelevant for the above argument; it only plays a role when we consider another Stokes transition across the negative real axis.
    \end{leftbar}
\end{exmp}
The structure of this example is a rather generic one for nonlinear problems, where the transmonomial $\re^{-nAz}$ is often multiplied by an equal power $\gs^n$ of a parameter $\sigma$ that can have the interpretation of a boundary condition or integration constant. The same structure occurs for example in solutions to ordinary differential equations (such as the Painlevé equations) or finite difference equations (such as string equations for matrix models) -- see \cite{Aniceto:2011nu} for these examples.

In order to deduce the precise map $g(\vec{\gs})$ that acts on the transseries parameters, and to prove equations such as \eqref{eq:BorelResurgenceGeneral} that describe how different nonperturbative sectors know about each other, one must turn to the \textit{alien calculus} developed by Écalle~\cite{ecalle1985fonctions}, which introduces a set of derivative operators $\gD_{\ell A}$ called {\em alien derivatives}. These operators -- see e.g.~\cite{sauzin2014introduction} for a formal definition -- act on formal power series and produce new formal power series. They are labeled by `jumps' $\ell A$ between the singularities in the Borel plane. A central result in the theory of resurgence states that one can express the Stokes automorphism in terms of these operators as follows:
\be
\mathfrak{S} =\exp\left(\sum_{\ell=1}^\infty \re^{-\ell A z}\gD_{\ell A} \right)\,.
\label{eq:StokesAutoDef}
\ee
This equation is the most important piece of machinery under the hood of resurgence, but for the purposes of this paper, we can work with Borel plane equations like \eqref{eq:BorelResurgenceGeneral} and will not need the subtleties of alien calculus.

\subsection{Multivalued Borel residues}
\label{sec:AmbStokesData}
As we mentioned in section \ref{sec:stokes}, the Borel residues $\fS_{k\to k+\ell}$ depend on the path along which we analytically continue the Borel transforms from one point to another. In computing the discontinuity along a Stokes line, (\ref{eq:discontinuity}), we assumed that the Hankel contours run off to infinity {\em below} the Stokes line, meaning that we could use the Borel residues $\fS_{k\to k+\ell}$ as defined by paths {\em above} the Stokes line -- the convention in the left of figure \ref{fig:residueconvention}. 
Alternatively, we could consider diagonal Hankel contours {\em above} the Stokes line, in which case we would need to use a different set of residues that we denote by $\tilde{\fS}_{k\to k+\ell}$ and which correspond to analytic continuation of the Borel transforms along paths {\em below} the Stokes line, as in figure \ref{fig:residueconvention} on the right. 

To relate the two sets of Borel residues, we observe that both choices should describe the same discontinuity:
\bea
    \disc\, \varphi^{(0)}  & = &  - \cS_{-}\left(\sum_{k=1}^\infty  \fS_{0\to k} \varphi^{(k)} \re^{-kAz}\right)  = -\cS_{-} \left(\mathfrak{S} -1\right)\varphi^{(0)} \ret
   & =& - \cS_{+}\left(\sum_{k=1}^\infty \tilde \fS_{0\to k} \varphi^{(k)} \re^{-kAz} \right) =  -\cS_{+} \left(1-\mathfrak{S}^{-1}\right)\varphi^{(0)} \,.
    \label{eq:disctwice}
\eea
The top (bottom) line in this equation is obtained by letting the Hankel contours run to infinity below (above) the real line, as shown by the diagrams in the top (bottom) of figure~\ref{fig:TwoHankels}. This equation~(\ref{eq:disctwice}) implies that two conventions lead to two {\em distinct} transseries expansions that after the appropriate lateral resummation should reproduce the {\em same} discontinuity. Hence, these respective transseries must consist of different linear combinations of the formal power series $\varphi^{(k)}$, and therefore they are defined by two distinct sets of residues $\{\fS_{k\to k'}\}$ and $\{\tilde \fS_{k \to k'}\}$. To reproduce the same discontinuity, the two transseries expressions should be related via
\be
\mathfrak{S} \,\left( \sum_{k=1}^\infty \tilde \fS_{0\to k} \varphi^{(k)}\re^{-kAz}\right) =\sum_{k=1}^\infty \fS_{0 \to k} \varphi^{(k)}\re^{-kAz}. 
\label{eq:StokesJumpDisc}
\ee
If we know one set of residues, we can derive the other set straightforwardly by using this equation and the Stokes automorphism \eqref{eq:StokesJump2}. Using the Stokes matrix $\mathbb{S}$ we can interpret the above equation as saying that
\be
    \mathbb{S} \cdot \begin{pmatrix}
        0\\
        \tilde{\fS}_{0\to 1} \\
        \tilde{\fS}_{0\to 2} \\
        \tilde{\fS}_{0\to 3} \\
        \dots
    \end{pmatrix} = \begin{pmatrix}
        0\\
        \fS_{0\to 1} \\
        \fS_{0\to 2} \\
        \fS_{0\to 3} \\
        \dots
    \end{pmatrix} \,
    \label{eq:StokesVecMult}
\ee
Solving this equation for the first three Borel residues $\tilde{\fS}_{0\to k}$, we find that
\bea
    \tilde \fS_{0\to1} & = & \fS_{0\to1} \ret
    \tilde \fS_{0\to2} & = & \fS_{0\to2}+\fS_{0\to1}\fS_{1\to2} \ret
    \tilde \fS_{0\to3} & = & \fS_{0\to3}+\fS_{0\to1}\fS_{1\to3}+\fS_{0\to2}\fS_{2\to3}+\fS_{0\to1}\fS_{1\to2}\fS_{2\to3}\, .
    \label{eq:StokesAmbRelation123}
\eea
The general solution is
\be
    \tilde \fS_{k\to k+\ell} = \sum_{r =1}^\ell\,\sum_{k=k_0< \ldots <k_r = k+\ell  }\, \prod_{j=0}^{r-1} \fS_{ k_{j}\to k_{j+1}}.
    \label{eq:StokesAmbRelation}
\ee
Note that all the `one-step' residues $\fS_{k\to k+1} = \tilde \fS_{k\to k+1}$ are equal. This is as expected, since their respective paths are in the same homotopy class (see figure \ref{fig:residueconvention}). Of course, for the `multi-step' residues, we could have chosen even more complicated homotopy classes for the paths, going above some singularities and below others, or even circling the singularities several times -- but the two conventions above are clearly the most natural ones. Regardless of the conventions that we pick, any complete set of residues contains the same information about the resurgence relations (\ref{eq:BorelResurgence}) as any other complete set. 
    \begin{exmp}
    \label{ex:alternativeBorelResiduesFree}
    \begin{leftbar}
        For the quartic free energy, whose Borel residues $\fS_{k\to k+l}$ were given in \eqref{eq:ResidueQFE1}, one can also calculate the alternative set of Borel residues using \eqref{eq:StokesAmbRelation}, leading to 
        \be
            \tilde \fS_{k \to k+l} =  (-1)^{l} \, \binom{k+l}{k}S_1^l\, .
            \label{eq:ResidueQFE2}
        \ee
        These are again expressed in powers of a single Stokes constant $S_1$.
    \end{leftbar}
    \end{exmp}

\subsection{Derivation of large order formulas}
\label{sec:classicallargeorder}
Now, let us use the material reviewed in the previous subsections to derive the standard form in which large order formulas usually appear, expressed in terms of Borel residues to keep the discussion as generic as possible. We first note that using Cauchy's theorem, the series coefficients can be extracted from the Borel transform in the following way:
\be
    \frac{\varphi_g^{(0)}}{g!} = \oint_0 \frac{\df t}{2\pi \ri} \frac{\cB[ \varphi^{(0)}](t)}{t^{g+1}}\, .
    \label{eq:firststepLOR}
\ee
We can then deform the contour around $t=0$ to a contour around infinity and in the process pick up various Hankel contours associated to singularities that the Borel transform has. Given our definition of the Stokes automorphism, each of these Hankel contours $H_{kA}^-$ grabs the $k$-th singularity located at $kA$ and `leaves' the Borel plane below the positive real axis -- as shown in figure~\ref{fig:TwoHankels}. Using the resurgence relations we can expand the Borel transform locally near each of these, yielding
\be
    \frac{\varphi_g^{(0)}}{g!} = \sum_{k\geq1} \mathsf{S}_{0\to k}\int_{H_{kA}^-} \frac{\df t}{(2\pi \ri)^2} \, \frac{\cB [\varphi^{(k)}](t-kA)}{ t^{g+1} }\log(t-kA)+ \oint_{\infty} \frac{\df t}{2\pi \ri} \frac{\cB[ \varphi^{(0)}](t)}{t^{g+1}} \,.
    \label{eq:deformLOR}
\ee
The last term on the right hand side constitutes a contribution coming from infinity that might be relevant, depending on the details of the Borel transform $\cB [\varphi^{(0)}]$ and the value of $g$. For the rest of this computation we will assume that its contribution vanishes and therefore ignore it -- but we will come back to this assumption in sections \ref{sec:ELOR} and \ref{sec:ELORdiscussion}. By an appropriate shift, the right hand side can now be rewritten as
\be
    \frac{\varphi_g^{(0)}}{g!} = -\sum_{k\geq1} \mathsf{S}_{0\to k}\int_0^{\infty-\ri \eps} \frac{\df t}{2\pi \ri} \, \frac{\cB [\varphi^{(k)}](t)}{ (t+kA)^{g+1} } 
    \label{eq:LORinGammas}
\ee
where $\eps$ is an infinitesimally small positive number, causing the contour to run below the singularities. We can then replace the Borel transforms by their Taylor series around $t=0$ and integrate term by term with respect to $t$, which produces the familiar asymptotic large order relations that read\footnote{To anticipate our discussion in section 4, where we will want to analytically extend these expressions to complex values of $g$, we already write $\Gam(g)$ rather than $(g-1)!$ on the left hand side of this equation.}
\be
    \frac{2\pi \ri\, \varphi_g^{(0)} A^g}{\Gamma(g)} \simeq -\sum_{k\geq1}\mathsf{S}_{0\to k} k^{-g} \sum_{h\geq0}\frac{\Gamma(g-h) }{\Gamma(g)} (k A)^h\varphi^{(k)}_h
    \label{eq:LORinGammas1}.
\ee
Note that this expression only makes sense in the large $g$ limit, given the fact that the $h \geq g$ terms diverge. This is the form of the large order formulas that one often encounters in the literature, and that we wrote in \eqref{eq:LORintro} -- there with the Borel residues $\mathsf{S}_{0\to k}$ written out in terms of a Stokes constant $S_1$ as in (\ref{eq:ResidueQFE2}). An important observation (already underlying e.g.\ \cite{Aniceto:2011nu}) is now that this formula itself has the form of a transseries: by writing
\be
 k^{-g} = e^{-g \log(k)}
\ee
we see that it has `instanton actions' $\log(2), \log(3), \log(4)$ etc. The nature of this transseries is slightly different from the ones we have discussed so far, though; in particular, the instanton actions are not all integer multiples of a finite set of `basis actions', and they are (by the nature of the logarithmic function) not single-valued. One may therefore wonder about the resummation of this transseries, whether it also has a Stokes automorphism, how its Stokes data are related to those of the original transseries, and so on. It is to those questions that we turn in the next section.

\section{Resurgence of the large order transseries}
\label{sec:resurgenceofLOR}
We finished the previous section with the usual form of the large order relation, connecting the coefficients in the perturbative sector of a transseries to the coefficients in its nonperturbative sectors. In particular, we observed that this large order relation can be written in a form which itself looks like a transseries, with instanton actions $\log(k)$ for $k\geq2$. In this section we study this transseries -- which we shall call the \textit{large order transseries} -- in more detail. In section~\ref{sec:StirlingTransform} we discuss the so-called Stirling transform, which relates the large order transseries to the \textit{original} transseries. We show in section~\ref{sec:ResurgenceStructureLOR} how this Stirling transformation helps in particular to relate resurgence properties of the original transseries to that of the large order transseries, and how it explains several properties that this new transseries has, including the presumed periodicity along the imaginary axis in the Borel plane. In section~\ref{sec:StokesAutoLOR}, we then describe the Stokes phenomenon of the large order transeries and explain how one can unambiguously resum this transseries. Finally, in section~\ref{sec:ELOR}, we describe how this transseries can be used to derive an {\em exact} large order relation.

\subsection{The Stirling transform}
\label{sec:StirlingTransform}
For each `instanton number' $k$ in the large order transseries \eqref{eq:LORinGammas1}, the ratios of gamma functions can be expanded in powers of $1/g$. This yields a formal power series in $1/g$ for each $k$, whose coefficients encode the perturbative fluctuations $\varphi^{(k)}_h$ around the $k$-th instanton sector. For $k=1$, the first few terms of this series read
\be
    -\mathsf{S}_{0\to 1}\left(\varphi^{(1)}_0+\frac{A \varphi^{(1)}_1}{g}+\frac{A \varphi^{(1)}_1+A^2\varphi^{(1)}_2}{g^2} + \frac{A \varphi^{(1)}_1+3A^2 \varphi^{(1)}_2 +A^3 \varphi^{(1)}_3}{g^3}+ \cO(g^{-4}) \right)\,.
    \label{eq:LORk=1}
\ee
Because the 1-instanton coefficients $\varphi^{(1)}_h$ grow factorially as $h$ becomes large, the above power series will generically be asymptotically divergent as well. In fact, this is the case for all $k$-sectors in the large order relation (\ref{eq:LORinGammas1}).

Working out a few more orders in (\ref{eq:LORk=1}), we find that the integer coefficients appearing in the series correspond to \textit{Stirling numbers of the second kind} \cite{NIST:DLMF}. The above series therefore constitutes what we will call a \textit{rescaled Stirling transform} of the first instanton series. Let us elaborate: a formal power series $\psi(z) = \sum_{n\geq0} a_n z^{-n}$ defines a sequence of coefficients $a_n$ whose Stirling transform is another sequence of coefficients $b_n$ defined by
\be
    b_n = \sum_{k=0}^n  \stir{n}{k} \, a_k\,,
    \label{eq:Stirdefbn}
\ee
where the integers $\stir{n}{k}$ are the Stirling numbers of the second kind; see appendix \ref{sec:AppBorelStirling} for a definition. We then say that the formal power series
\be
    \tilde \psi(g) = \sum_{n=0}^\infty b_n g^{-n}
    \label{eq:Stirdefg}
\ee
is the Stirling transform of the series $\psi(z)$. Furthermore, we can rescale the argument $z \to z/A$ in $\psi(z)$ to obtain a new series in $z^{-1}$:
\be
    \psi\left(\frac{z}{A}\right) = \sum_{n=0}^\infty \left(a_n A^n\right) z^{-n}\,.
\ee
Finally, we introduce a formal power series $\hat \psi(g)$ which is the Stirling transform of this rescaled series, and which looks as follows:
\bea
    \hat \psi(g) & = & \sum_{n=0}^\infty \sum_{k=0}^n \stir{n}{k} \left(a_k A^k\right) g^{-n} \ret
    & = & a_0+\frac{A a_1}{g}+\frac{A a_1+A^2a_2}{g^2} + \frac{A a_1+3A^2 a_2 +A^3 a_3}{g^3}+ \cO(g^{-4}) \,.
    \label{eq:Stirdefgtil}
\eea
This series is what we call the rescaled Stirling transform of $\psi(z)$ and it matches exactly the series expansion \eqref{eq:LORk=1} of the large order relation \eqref{eq:LORinGammas1}. If we now identify the coefficients $a_h$ with the coefficients $\varphi^{(1)}_h$ of the one-instanton sector of our original transseries, then we conclude that the asymptotic expansion for $k=1$ in (\ref{eq:LORinGammas1}) is the rescaled Stirling transform of that one-instanton series. More generally, the asymptotic expansion in $1/g$ of the large order transseries in its nonperturbative sector multiplying $k^{-g} = e^{-g \log k}$ can be expressed as a rescaled Stirling transform of the $k$-instanton series of the original transseries, where the scaling factor is $k A$ instead of~$A$.

Having understood that the asymptotic expansions appearing in the large order relations are actually rescaled Stirling transforms of the original instanton series, we can define
\be
 \label{eq:Stirlingseries}
    \hat \varphi^{(k)}(g)
    = \sum_{n=0}^\infty \sum_{\ell=0}^n \stir{n}{\ell} \left(\varphi^{(k)}_\ell \left(k A \right)^\ell \right) g^{-n}\,,
\ee
and rewrite the asymptotic large order relations (\ref{eq:LORinGammas1}) as
\be
    \frac{2\pi \ri \varphi^{(0)}_g A^g}{\Gamma(g)} \simeq - \sum_{k=1}^\infty \mathsf{S}_{0\to k} \hat \varphi^{(k)}(g) \, \re^{-g \log(k)}\,.
    \label{eq:LORtransseries1}
\ee
Thus, \eqref{eq:LORtransseries1} allows us to order the contributions by size where the one-instanton coefficients $\varphi_h^{(1)}$ determine the leading expansion in $1/g$, and higher instanton coefficients $\varphi_h^{(k)}$ determine nonperturbative contributions to the large order transseries that are exponentially suppressed as $\re^{-g\log(k)}$. 

\begin{figure}[ht]
 \includegraphics[width= \linewidth]{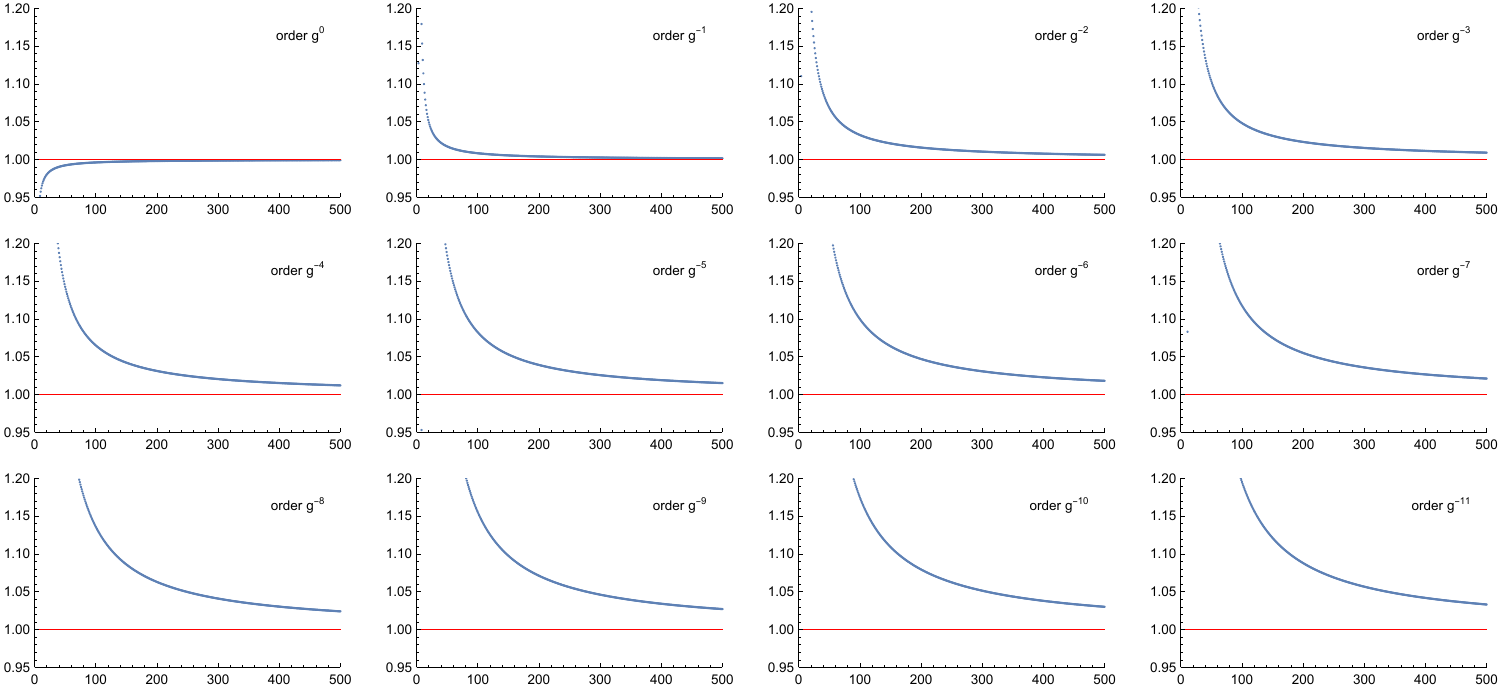}
 \caption{A test of the rescaled Stirling transform appearing in \eqref{eq:LORtransseries1}, for the leading $k=1$ sector and up to order $g^{-12}$.}
 \label{fig:testLO1}
\end{figure}

\begin{exmp}
\label{ex:ratioTests}
\begin{leftbar}
Let us test our expression (\ref{eq:LORtransseries1}) by performing a resurgent large order analysis at leading instanton order $k=1$. (We will perform tests for higher $k$ below in example \ref{ex:ExactResummingFreeEnergy}.) 
To this end (see also~\cite{Aniceto_2019}), we study the large order growth of perturbative coefficients $F^{(0)}_g$ of the quartic free energy, which we already briefly introduced in example \ref{ex:StokesAutoFreeEnergy}.

The large order relation for the quartic free energy is exactly of the form \eqref{eq:LORtransseries1}, with instanton sectors $F^{(k)}$ that carry an instanton action $A=3/2$ and with Borel residues $\mathsf{S}_{0\to k}$ expressed in terms of a Stokes constant $S_1=-2$ by \eqref{eq:ResidueQFE1}. To test~\eqref{eq:LORtransseries1}, we call the sequence on the left hand side of this equation $r_g$ and plot the ratio
\begin{equation}
\frac{r_g}{S_1 \hat{\varphi}_0^{(1)}}
\simeq 1+\cO(g^{-1})\,,
\end{equation}
for $g$ up to 500 in the first plot of figure~\ref{fig:testLO1}. This indeed convergences towards 1 as $g$ becomes large. Likewise, in the second plot, we test the expected subleading $g^{-1}$-behaviour by plotting the ratio 
\begin{equation}
g\frac{r_g-S_1\hat{\varphi}_0^{(1)}}{S_1\hat{\varphi}_1^{(1)}}\simeq 1+\cO(g^{-1})\,,
\end{equation}
which again converges towards 1 in the large $g$ limit. We can similarly test the $g^{-N}$ contributions to the leading sector of (\ref{eq:LORtransseries1}) by summing the right hand side of \eqref{eq:Stirlingseries} (for $k=1$) up to order $n=N-1$. We then plot the difference between the (exact) left and (truncated) right hand side of \eqref{eq:Stirlingseries} for values of $g$ up to 500, where we divide by the expected $g^{-N}$ contribution at that order so that we expect the rescaled difference to become 1 at large $g$. In figure \ref{fig:testLO1} we see that indeed both sides of~\eqref{eq:LORtransseries1} match up to order $g^{-12}$ (where the only reason to stop is not to make the figure too large), confirming the validity of the Stirling numbers appearing in~\eqref{eq:Stirlingseries}.
\end{leftbar}
\end{exmp}

At this point, we have established how the large order transseries \eqref{eq:LORtransseries1} is constructed in terms of the instanton coefficients $\varphi_h^{(k)}$ of the original transseries. Still, the question remains whether the newly constructed large order transseries is truly a \textit{resurgent} transseries, in the sense that the Borel transforms of its formal power series $\hat \varphi^{(k)}(g)$ know about each other like the original series $\varphi^{(k)}(g)$ do in~\eqref{eq:BorelResurgenceGeneral}. In order to answer this question, we want to derive similar resurgence relations for the Borel transforms of the rescaled Stirling transforms $\hat\varphi^{(k)}$. The first question to address is therefore what the Borel planes of these series look like.

It so happens that the Stirling transforms of functions have a rather interesting property when it comes to their Borel transforms: consider the Stirling transform $\tilde{\psi}(g)$  -- defined through (\ref{eq:Stirdefbn}) and (\ref{eq:Stirdefg}) from the series $\psi(z)$. Its 0-Borel transform \eqref{eq:nboreldef}, assuming as usual that it converges and through analytic continuation defines a function, has the following simple form:
\be
\label{eq:StirlingTrafoBorel0}
    \cB [\,\tilde \psi\,](t) = \cB[\,\psi\,](\re^t-1)\,.
\ee
Its generic $n$-Borel transform is straightforwardly obtained through differentiation, see appendix~\ref{sec:nBorel}. Subsequently, the rescaled Stirling transform $\hat \psi(g)$ given by (\ref{eq:Stirdefgtil}), also has a Borel transform that relates to $\cB[\,\psi\,]$ as follows:
\be
\label{eq:RescaledStirlingTrafoBorel0}
    \cB[\,\hat \psi \,](t) = \cB[\,\psi\,]\left(A(\re^t-1)\right)\,.
\ee
As a result, we find that the 0-Borel transforms of the asymptotic series $\hat \varphi^{(k)}(g)$ appearing in the large order relations can be expressed explicitly in terms of the Borel transforms of the original instanton series $\varphi^{(k)}$, using their proper rescaling:
\be
    \cB [\hat \varphi^{(k)}](t) =  \cB [ \varphi^{(k)}]\left(kA(\re^t-1)\right)
    \label{eq:StirlingTrafoBorel1}\,.
\ee
We provide derivations of these statements in appendix \ref{sec:AppBorelStirling}. The above relations already reveal the underlying reason for the appearance of the towers of singularities in the Borel plane that we discussed in the introduction: the Borel transforms \eqref{eq:StirlingTrafoBorel1} of the rescaled Stirling series are functions of $\re^t$ and are therefore invariant under $t \to t+2\pi \ri$. As a result, any singularity in the Borel plane of the original instanton series $\cB [\varphi^{(k)}]$ reappears as a tower of evenly spaced singularities in the Borel plane of $ \cB [\hat\varphi^{(k)}]$. As we explain in the next subsection, this `translational symmetry' also explains why it makes sense to write $k^{-g}$ as $\re^{-g \log(k)}$ in terms of the `multivalued instanton action' $\log(k)$ in the large order transseries~(\ref{eq:LORtransseries1}).\footnote{In a different context, the change of variables $s = kA(\re^t-1)$ is the \textit{uniformizing map} in \cite{Costin:2021bay} for a Borel transform with a single singularity at $s=-kA$.}

\subsection{Resurgent structure}
\label{sec:ResurgenceStructureLOR}
Now that we have seen that the Borel transforms of the rescaled Stirling transforms of the $\varphi^{(k)}(z)$ can be expressed in terms of the Borel transforms of these instanton series themselves, we can derive resurgence relations like (\ref{eq:BorelResurgenceGeneral}) for the rescaled Stirling transforms. Using \eqref{eq:StirlingTrafoBorel1} we obtain
\bea
    \cB[\hat\varphi^{(k)}](t) & = & \cB[\varphi^{(k)}]\left(kA(\re^t-1)\right) \label{eq:ResRelForStirling}\\
    &=& \frac{\fS_{k\to k+\ell}}{2\pi \ri} \cB[\varphi^{(k+\ell)}]\left(kA(\re^t-1)-\ell A\right)\log\left(kA(\re^t-1)-\ell A\right) + \text{regular} \ret
    & = & \frac{\fS_{k\to k+\ell}}{2\pi \ri}\cB[\hat \varphi^{(k+\ell)}]\left(t-\log\left(\frac{k+\ell}{k}\right)\right)\log\left(t-\log\left(\frac{k+\ell}{k}\right)\right)+\text{regular}\,,\nonumber
\eea
where in going from the first to the second line we expanded $s \equiv kA(\re^t-1)$ around $s =\ell A$ using the ordinary resurgence relations (\ref{eq:BorelResurgenceGeneral}). In the last line we expanded the logarithm around $s =\ell A $ and used \eqref{eq:StirlingTrafoBorel1} to go back to the Borel transform of $\hat{\varphi}^{(k+\ell)}$. Notice that we are just interested in the singular terms in these expressions; the higher order terms in the expansion of the logarithm only yield regular terms, since they multiply a Borel transform which is analytic at $s=\ell A$.

The resurgence relations for the transseries sectors $\hat \varphi^{(k)}$ of the large order transseries,~\eqref{eq:ResRelForStirling}, are quite similar to \eqref{eq:BorelResurgenceGeneral} describing the resurgence of the original transseries sectors $\varphi^{(k)}$. We observe in particular that the large order transseries sectors labeled by $k$ and $k+\ell$ are connected by the same Borel residues~$\fS_{k\to k+\ell}$ that occurred for the original transseries. An important difference lies in the fact that the locations of the Borel singularities are completely different from those of the original transseries. Within the original transseries, the $k$-instanton sector knows about the $(k+\ell)$-sector by expanding $\cB[\varphi^{(k)}](t)$ around $t = \ell A$. In the case of our large order transseries, the $k$-sector knows about the $(k+\ell)$-sector by expanding $\cB[\hat \varphi^{(k)}](t)$ around $t = \log\left(\frac{k+\ell}{k}\right)$. In fact, due to the $t \sim t+2\pi \ri$ symmetry of \eqref{eq:StirlingTrafoBorel1}, we can add any integer multiple of $2\pi \ri$ to the location of a given singularity of $\cB[\hat \varphi^{(k)}](t)$ to find another singularity from which the same $(k+\ell)$-sector resurges with the same Borel residue~$\fS_{k\to k+\ell}$. 

Summarizing this discussion, the Borel transform $\cB[\hat \varphi^{(k)}](t)$ will have singularities at 
\be
    \cA_{\ell, n}^{(k)} = \log\left(\frac{k+\ell}{k}\right)+2\pi \ri n \qquad \text{for } k+\ell > 0 \text{ , } \ell\neq0 \text{ and } n\in \mathbb{Z}\, ,
    \label{eq:LORdefA}
\ee
from which the $(k+\ell)$-sector resurges with residue $\fS_{k\to k+\ell}$. When we lift the condition that $k+\ell>0$, which happens for example in the case with two opposite primitive instanton actions $A_1=-A_2$, we can also get singularities at 
\begin{equation}\label{eq:LORdefA2}
\tilde{\cA}_{\ell, n}^{(k)} 
= \log\left(-\frac{k+\ell}{k}\right)+ (2n+1) \pi \ri \qquad \text{for } k+\ell < 0 \text{ , } \ell\neq0 \text{ and } n\in \mathbb{Z}\,.
\end{equation}
This pattern of singularities is for example relevant for the plots shown in figure~\ref{fig:BPplotAdler} in the introduction. One can write similar formulas for more general cases (which we will not consider further in this paper) when there are several primitive instanton actions $A_1, A_2, \ldots$, even when there is no relation between them, where the addition of $\pi \ri$ in \eqref{eq:LORdefA2} changes.

Using the above notation, the resurgence relations in \eqref{eq:ResRelForStirling} can be written compactly as
\be
\cB[\hat\varphi^{(k)}](t)\bigg|_{t=\cA^{(k)}_{\ell,n}}
= \frac{\mathsf{S}_{k\to k+\ell}}{2 \pi i} \  \cB[\hat \varphi^{(k+\ell)}]\left(t-\cA^{(k)}_{\ell,n}\right) \ \log\left(t-\cA^{(k)}_{\ell,n}\right)+\text{regular}\,,
\label{eq:BorelResurgenceLOR}
\ee
which also makes the similarity to \eqref{eq:BorelResurgenceGeneral} more manifest. As for `ordinary' resurgence, these large order resurgence relations tell us that when we integrate around any singularity $\cA^{(k)}_{\ell,n}$ in the Borel plane, we can extract the formal power series $\hat \varphi^{(k+\ell)}$. In the spirit of section \ref{sec:backrgound}, this suggests that we should unite all these series into what we call the \textit{generic large order transseries}
\begin{equation}
\Phi(\gs_{k,n},g)
= \sum_{n=-\infty}^\infty \sum_{k=1}^\infty\gs_{k,n}\ \hat \varphi^{(k)}(g) \, \re^{-g\cA^{(1)}_{k,n}}=\sum_{n=-\infty}^\infty \Phi_n(\vec{\gs}_n, g)\,.
\label{eq:LORtransseries2}
\end{equation}
The parameters $\gs_{k,n}$ are again transseries parameters associated to sectors $(k, n)$. In section~\ref{sec:StokesAutoLOR} we will see how the Stokes automorphism acts on these parameters. In anticipation of what is to come, it is also useful to write the \textit{full transseries} $\Phi$ as above, as an infinite sum over `horizontal' \textit{subtransseries} $\Phi_n$. The large order transseries \eqref{eq:LORtransseries1} is then obtained by setting $\gs_{k,n} = -\fS_{0\to k} \delta_{n,0}$. Note that the locations $\cA_{\ell,n}^{(k)}$ of the Borel singularities in the $k$-th Borel plane correspond to the relative {\em distances} between the instanton actions of the sectors $(k+\ell, m+n)$ and $(k, m)$ in the large order transeries:
\be
    \cA^{(k)}_{\ell,n}  = \cA^{(1)}_{k+\ell,m+n}-\cA^{(1)}_{k,m}\, ,
\ee
for arbitrary integers $m$. These distances are visible in the Borel plane of the leading sector -- see also to the singularities shown in figure \ref{fig:BPplotAdler}. The fact that the singularities of the Borel transforms correspond to relative distances between instanton actions holds in general, but is often concealed by the fact that in ordinary transseries these actions lie on a semi-lattice generated by a single primitive instanton action $A$. For the large order transseries, on the other hand, there are infinitely many primitive instanton actions $\log(p)$ for $p$ prime.

We conclude that the large order transseries is truly a resurgent transseries, just like the original transseries. The only differences are that its singularities no longer depend on the instanton action $A$ of the original transseries, and that each singularity associated to a sector $k+\ell$ is copied at locations shifted by integer multiples of $2\pi \ri$ in the Borel plane of $\cB[\hat \varphi^{(k)}]$. The domain of the variable $t$ in terms of which the Borel transforms $\cB[\hat \varphi^{(k)}]$ are defined is thus effectively a cylinder that we will henceforth call the \textit{Borel cylinder}. This cylinder provides the best way to structure the singularities found in the figures~\ref{fig:BPplotAdler} and~\ref{fig:BPplotFreeEnergy} that we encountered in section~\ref{sec:intro}; we discuss this further in example \ref{ex:AdlerCylinder} below.

\begin{figure}
\centering
\subfloat[]{
    \includegraphics[width=.46\textwidth]{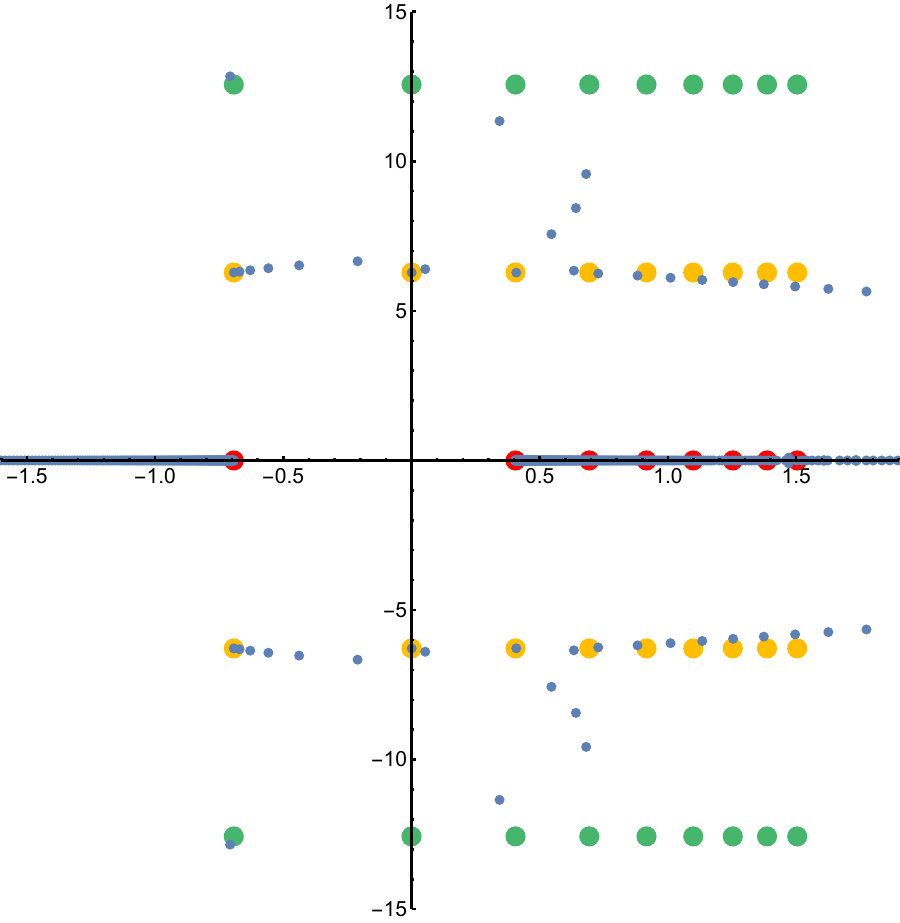}
    \label{fig:BP_plot_free_energy2}
}
\subfloat[]{
    \includegraphics[width=.48\textwidth]{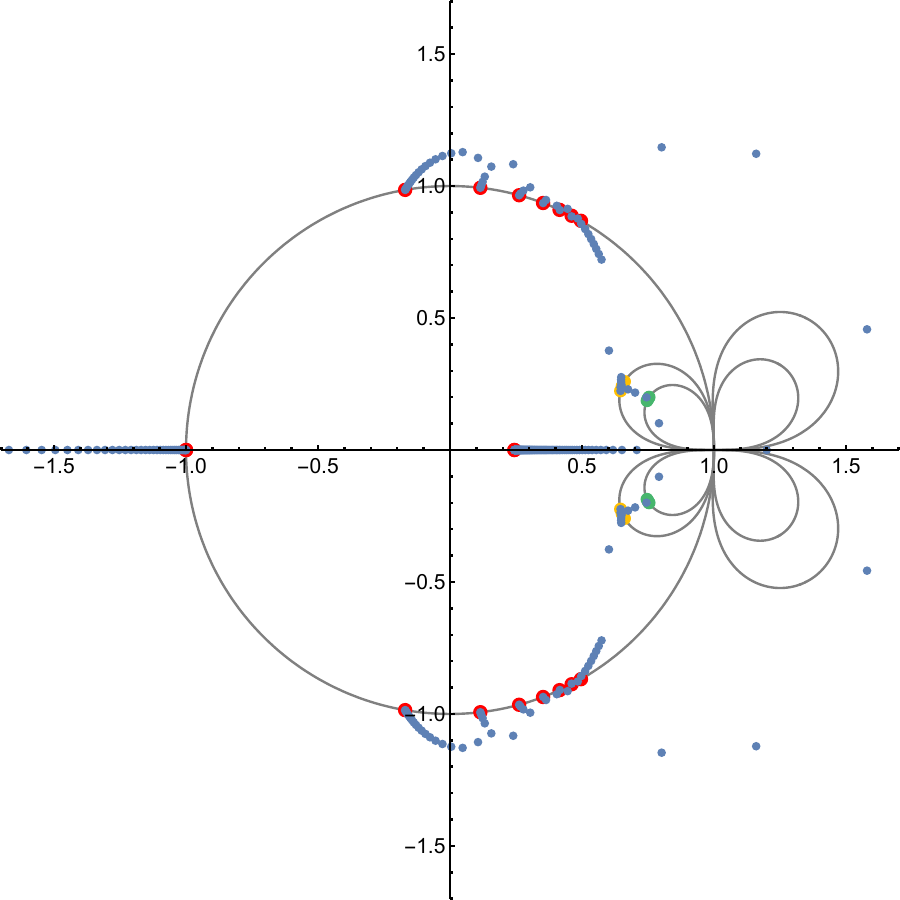}
    \label{fig:CBP_plot_free_energy2}
}
\caption{In figure~\ref{fig:BP_plot_free_energy2}, we show the singularities of the diagonal order $800$ Padé approximant of $\cB[\hat{F}^{(2)}](t)$. As the branch cuts are mimicked by an accumulation of poles, the poles corresponding to different branch cuts on the real line are hard to distinguish. We therefore show in figure~\ref{fig:CBP_plot_free_energy2} the order $300$ diagonal Padé approximant of $\cB[\hat{F}^{(2)}]$ after applying the conformal map of \eqref{eq:conformalMap}. The branch cuts positioned on the positive real line are now pulled apart as they are mapped to the unit circle. The singularity at $t=\log(1/2)=-\log(2)$ of figure~\ref{fig:BP_plot_free_energy2} gets mapped to $z\approx 0.24$. Furthermore, the `flower-like' curves in the $z$-plane correspond to $\Im(t)=\pm2\pi\ri$ and $\Im(t)=\pm4\pi\ri$. Unfortunately, the branch cut starting at $z\approx 0.24$ `uses' many poles of the Padé approximant, so we only observe a limited number of singularities starting on these curves. As a result, the branch cuts at $\Im(t)\pm2\pi n\ri$ for $n \neq 0$ are instead best viewed in the original $t$-plane.}
\label{fig:BPplotFreeEnergy2}
\end{figure}

\begin{exmp}
\label{ex:AdlerCylinder}
\begin{leftbar}
Let us see whether in examples we can reproduce the singularity structure $\cA^{(k)}_{\ell,n}$ and $\tilde \cA^{(k)}_{\ell,n}$ that is described in \eqref{eq:LORdefA} and \eqref{eq:LORdefA2}. Since we do not have closed form expressions for the various Borel transforms, we make use of Padé approximants (see e.g.\ \cite{bender78:AMM}) to mimic their singularity structure. The singularities $\cA^{(1)}_{\ell, n}$ and $\tilde \cA^{(1)}_{\ell,n}$, as approximated by the Padé approximant of $\cB[\hat \varphi^{(1)}]$, were already shown in figures~\ref{fig:BPplotAdler} and~\ref{fig:BPplotFreeEnergy} in the introduction, for the Adler function and the quartic free energy respectively. As further evidence for the structure, we also consider the second rescaled Stirling transform for the quartic free energy. In figure~\ref{fig:BPplotFreeEnergy2} we plot the singularities of the Padé approximant of the Borel transform $\cB[\hat{F}^{(2)}]$ of this series.

Starting with figure~\ref{fig:BPplotAdler1} for the Adler function, we clearly observe singularities at the anticipated positions $t=\cA_{\ell,n}^{(1)}$ with $\ell=1,2$. 
For figure~\ref{fig:BPplotAdler2}, besides singularities at $t=\cA_{\ell,n}^{(1)}$ with $\ell=2$, we also notice singularities at positions $t=\tilde{\cA}_{\ell,n}^{(1)}$ , with $\ell=1,2$. This is a consequence of the fact that the Adler function has two opposite primitive instanton actions $A_1=-A_2$, which as we have explained leads to additional contributions in the large order relation \eqref{eq:LORtransseries1} that scale with $(-k)^{-g}=e^{-g\log(k)\pm\pi\ri}$, leading in turn to \eqref{eq:LORdefA2}.

For figures~\ref{fig:BP_plot_free_energy} and~\ref{fig:BP_plot_free_energy2} we also observe singularities at $t=\cA_{\ell,n}^{(1)}$. However, as the branch cuts starting at $t=\log(k)$, $k\geq2$ are mimicked by the accumulation of Padé poles along the real axis, it is hard to distinguish the branch cuts that start at $t=\log(k)$. We have therefore applied a conformal mapping~\cite{Costin:2020hwg,Costin:2021bay}
\begin{equation}
\label{eq:conformalMap}
t=-\frac{4z\log\left(\frac{k+1}{k}\right)}{(1-z)^2}\,,
\end{equation}
which maps the half-line $[\log\frac{k+1}{k},\infty)$ to the unit circle in the $z$-plane, where in particular $t=\log\frac{k+1}{k}$ gets mapped to $z=-1$. The resulting figures for the Padé approximants of these conformally mapped series are shown in figures~\ref{fig:CBP_plot_free_energy} and~\ref{fig:CBP_plot_free_energy2}. We observe that the branch cuts starting on the real positive $t$-axis are now pulled apart in the $z$-plane and for low values of $k$ become clearly visible. 
\end{leftbar}
\end{exmp}

\subsection{Stokes automorphism}
\label{sec:StokesAutoLOR}
In the previous section, we have verified that the large order transseries \eqref{eq:LORtransseries1} and its generic version \eqref{eq:LORtransseries2} truly form {\em resurgent} transseries by establishing the resurgent structure \eqref{eq:BorelResurgenceLOR} of their nonperturbative sectors. For practical purposes, we are often interested in resumming the sectors $\hat \varphi^{(k)}$ that appear in these large order transseries. The reason for this (as explained elaborately in e.g.\ \cite{Aniceto_2019}), is that we often want to use the large order behaviour of the perturbative coefficients $\varphi^{(0)}_g$ -- which may be easy to obtain in specific examples --  to decode the nonperturbative coefficients $\varphi^{(k)}_h$ of the $k$-instanton sector, which may not be so easy to obtain by other methods, for example because the nonperturbative effects in the underlying physical problem are not {\em a priori} known. To achieve such a decoding, we need to first `remove' all sectors $\hat \varphi^{(k')}$ for which $k' <k$ from the large order transseries before we can read off the coefficients in the $k$-th sector. The only effective way to do this is by resumming these series $\hat \varphi^{(k')}$ and substracting them from the left hand side of the large order transseries. As is so often the case, to achieve this we need to decide along which contour in the Borel plane -- or here: the Borel cylinder -- we resum, and different choices are related by a Stokes phenomenon. In this subsection we discuss various contours and use the resurgence relations \eqref{eq:BorelResurgenceLOR} as well as consistency of resummation to derive the Stokes phenomenon of the generic large order transseries.
\begin{figure}
    \centering
    \includegraphics[width = 0.8\linewidth]{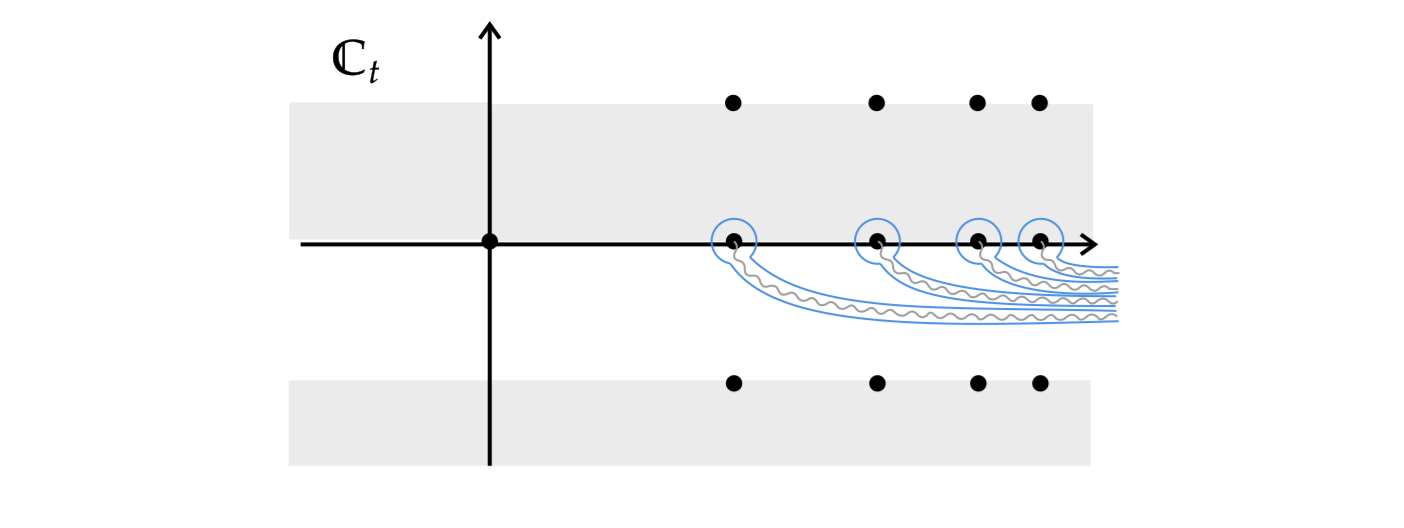}
    \caption{The Hankel contours $H^-_{\log(1+\ell)}$ along which we integrate in equation \eqref{eq:discLORsector1}.}
    \label{fig:HankelLOR}
\end{figure}

As we saw in \eqref{eq:LORdefA}, the singularities of the large order transseries sectors $\hat \varphi^{(k)}$ accumulate along horizontal lines separated by integer multiples of $2\pi \ri$. There are two natural resummations $\cS_{0^+}$ and $\cS_{0^-}$, whose contours go above and below the singularities on the real positive line respectively:
\be
    \cS_{0^\pm} \hat \varphi^{(k)}(g) = g \int_0^{\infty\pm \ri \eps} \cB[\hat \varphi^{(k)}](t) \, \re^{-gt} \df t \, .
    \label{eq:resumLORdef0}
\ee
We can then define the discontinuity operator
\be
    \disc_0 = \cS_{0^+}-\cS_{0^-} \, ,
\ee
in order to study the Stokes phenomenon as discussed in section \ref{sec:stokes}. The discontinuity of the first large order transseries sector $\hat \varphi^{(1)}$ can readily be computed, and yields
\bea
    \disc_0 \,\hat\varphi^{(1)} & = &\sum_{\ell=1}^\infty \mathsf{S}_{1\to 1+\ell} \ g \int_{H_{\log(1+\ell)}^-} \frac{\df t}{2\pi \ri}\, \cB [\hat \varphi^{(1+\ell)}](t-\log(1+\ell))\log\left( t-\log(1+\ell)\right) \re^{-gt} \ret
    &=& -\sum_{\ell=1}^\infty \mathsf{S}_{1\to 1+\ell}\, \re^{-g \log(1+\ell)} \cS_{0^-} \hat \varphi^{(1+\ell)}\,,
    \label{eq:discLORsector1}
\eea
where the Hankel contours $H_{\log(1+\ell)}^-$ are shown in figure \ref{fig:HankelLOR}. We can repeat this computation for generic sectors $\hat \varphi^{(k)}$ and find that 
\be
\disc_0 \, \hat\varphi^{(k)} = -\sum_{\ell=1}^\infty \mathsf{S}_{k\to k+\ell} \, \re^{-g \log\left(\frac{k+\ell}{k}\right)} \cS_{0^-} \hat \varphi^{(k+\ell)}\,.
\ee
Note that the instanton actions that appear correctly correspond to the singularities $\cA^{(k)}_{l,0}$ in \eqref{eq:LORdefA}. From the discontinuity of the large order transseries sectors $\hat \varphi^{(k)}$, we extract the Stokes phenomenon
\be
    \mathfrak{S}_0 \hat \varphi^{(k)} = \hat \varphi^{(k)}- \sum_{\ell=1}^\infty \mathsf{S}_{k\to k+\ell} \,  \hat \varphi^{(k+\ell)}\,\re^{-g \log\left(\frac{k+\ell}{k}\right)} \,,
    \label{eq:StokesJumpLOR1}
\ee
which is the direct analogue of \eqref{eq:StokesJump2} for the original transseries. The action of the Stokes automorphism can be formulated within the horizontal subtransseries $\Phi_n(g, \vec{\gs}_n)$ of \eqref{eq:LORtransseries2} as follows:
\be
    \mathfrak{S}_0\Phi_n\left(g,\vec{\gs}_{n}\right)= \Phi_n\left(g,\mathbb{S}_0 \cdot \vec{\gs}_{n}\right)  \, , 
    \label{eq:StokesJumpLOR2}
\ee
or at the level of individual transseries parameters as
\be
    \mS_0: \gs_{k,n} \mapsto \gs_{k,n}- \sum_{j=1}^{k-1} \gs_{j,n} \fS_{j \to k} \, .
    \label{eq:LORStokesRow}
\ee
This mapping should be compared to \eqref{eq:Stokesactionpar} for the original transseries. The matrix $\mathbb{S}_0$ is essentially the same as $\mathbb{S}$ given in \eqref{eq:StokesMatrix}, but with the `zeroth' row and column removed. Since a $\hat \varphi^{(0)}$ sector does not exist in the large order transseries, we can assume such a sector to vanish identically and naturally extend the Stokes matrix $\mathbb{S}_0$ to include also a $j=0$ contribution in \eqref{eq:LORStokesRow}, in which case we simply have 
\be
    \mathbb{S}_0 = \mathbb{S}\,.
    \label{eq:equivalentStokesMatrix}
\ee
The equivalence of the Stokes matrix associated to the Stokes phenomenon of the large order transseries across the positive real axis, to that of the ordinary transseries, also means the following. If we consider our large order transeries \eqref{eq:LORtransseries1} and study its Stokes phenomenon in the \textit{anti-clockwise} direction, given by the inverse automorphism $\mathfrak{S}_0^{-1}$ or equivalently by the inverse Stokes matrix $ \mathbb{S}^{-1}$, we retrieve the alternative Borel residues by virtue of \eqref{eq:StokesVecMult}:

\be
    \mathbb{S}^{-1} \cdot \begin{pmatrix}
        0\\
        \fS_{0\to 1} \\
        \fS_{0\to 2} \\
        \fS_{0\to 3} \\
        \dots
    \end{pmatrix} = \begin{pmatrix}
        0\\
        \tilde{\fS}_{0\to 1} \\
        \tilde{\fS}_{0\to 2} \\
        \tilde{\fS}_{0\to 3} \\
        \dots
    \end{pmatrix} \, .
\ee
Said otherwise, we have that
\be
    \mathfrak{S}_0 \left( \sum_{k=1}^\infty \tilde{\mathsf{S}}_{0\to k} \hat \varphi^{(k)}(g) \re^{-g\log(k)}\right) = \sum_{k=1}^\infty \mathsf{S}_{0\to k} \hat \varphi^{(k)}(g) \re^{-g\log(k)}\, .
    \label{StokesAutoLOR}
\ee
That is, the large order transseries with transseries parameters $\tilde \fS_{0\to k}$ resummed \textit{above} the positive real line (with $\cS_{0^+}$) is equivalent to the large order transseries with parameters $\fS_{0\to k}$ resummed \textit{below} the positive real line (with $\cS_{0^-}$). This equation is the direct analogue of \eqref{eq:StokesJumpDisc} for the original transseries.

\begin{figure}
    \centering
    \includegraphics[width = 0.9\linewidth]{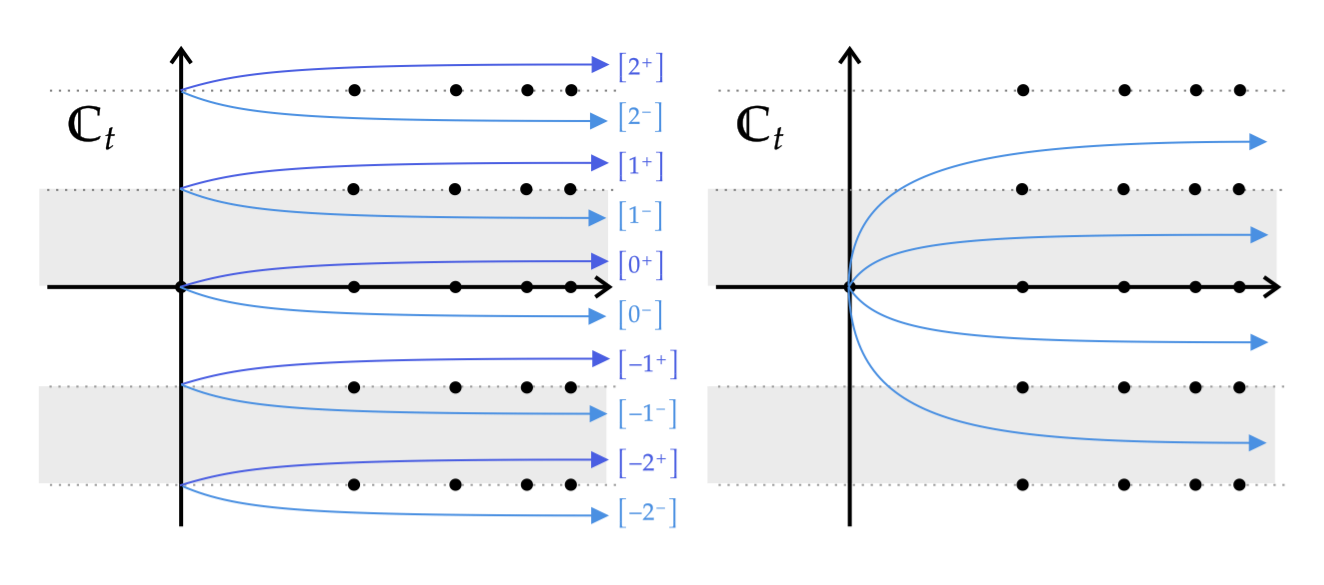}
    \caption{On the left: the contours $[n^\pm]$ corresponding to the resummations $\cS_{n^\pm}$ defined in \eqref{eq:discLORsector1}. Once we have derived the Stokes phenomena $\mS_{n}$ across the $n$th horizontal ray of singularities, we are able to resum the large order transseries along any of the contours shown on the right. }
    \label{fig:contourcylinder}
\end{figure}

We can generalize the discontinuity $\disc_0$, across singularitities that lie on the positive real line of the Borel plane, to Stokes transitions with discontinuities that we will label $\disc_n$ and that arise from singularities $\cA^{(k)}_{l, n}$ on other horizontal lines with fixed $n$ -- see figure \ref{fig:contourcylinder}. For completeness, let us derive the Stokes automorphisms for these transitions and see how they alter the transeries parameters $\gs_{k,n}$.  

Consider the contours shown in figure \ref{fig:contourcylinder} on the left, which define Borel summations
\be
    \cS_{n ^\pm} \hat \varphi^{(k)}(g) = g \int_{2\pi \ri n}^{\infty +2\pi \ri n \pm \ri \eps} \cB[\hat \varphi^{(k)}](t) \, \re^{-gt} \df t \, .
    \label{eq:discLORsector2}
\ee
By the symmetry of the Borel cylinder, these resummations can be related to \eqref{eq:resumLORdef0} as follows:
\be
    \cS_{n^\pm}\hat \varphi^{(k)}(g) = \re^{-2\pi \ri n g} \cS_{0^\pm}\hat \varphi^{(k)}(g)\,.
    \label{eq:ResumNLOR}
\ee
For integer $g$, the prefactor on the right hand side equals 1, reflecting the symmetry of the Borel cylinder under translations by $2 \pi \ri$. In what follows, we will also be interested in cases with non-integer $g$ (where for real $g$ we can mathematically think of the factors of $\re^{-2\pi \ri n g}$ as transition functions on a nontrivial $U(1)$-bundle over the Borel cylinder), so we will keep these factors explicit in our formulas. We can now define discontinuities $\disc_n = \cS_{ n ^+} -\cS_{n ^-}$, which can be expressed as
\be
\label{eq:discLORtransseries}
    \disc_n \, \hat \varphi^{(k)}(g) = \re^{-2\pi\ri n g}\,\disc_0\,  \hat \varphi^{(k)}(g) \, .
\ee
In terms of Stokes automorphisms we then have
\be
    \mathfrak{S}_n \hat \varphi^{(k)}(g) = \hat \varphi^{(k)}(g) - \re^{-2\pi \ri n g} \sum_{\ell=1}^\infty \mathsf{S}_{k\to k+\ell} \,  \hat \varphi^{(k+\ell)}(g)\,\re^{-g \log\left(\frac{k+\ell}{k}\right)}\,.
    \label{eq:StokesJumpLOR3}
\ee
We note that the sectors $\varphi^{(k+\ell)}$ and the Stokes residues $\mathsf{S}_{k\to k+\ell}$ appear, as was to be expected, but with an instanton action that is shifted by $-2\pi \ri n$. 

Next, we can derive the action of this Stokes automorphism on the subtransseries of~\eqref{eq:LORtransseries1}:
\be
     \mathfrak{S}_n \Phi_m(\vec{\gs}_m, g) = \Phi_{m}(\vec{\gs}_m, g)+\Phi_{m+n}\left(\left(\mathbb{S}-\mathrm{I}\right)\cdot\vec{\gs}_m, g\right)\, ,
     \label{eq:StokesJumpLOR4}
\ee
where $\mathrm{I}$ is an infinite identity matrix. As a consistency check, note that \eqref{eq:StokesJumpLOR3} and  \eqref{eq:StokesJumpLOR4} reduce to \eqref{eq:StokesJumpLOR1} and \eqref{eq:StokesJumpLOR2} respectively when we set $n=0$. For the full transseries, we find that 
\be
    \mathfrak{S}_n \Phi(\gs, g) = \sum_{m=-\infty}^\infty \Phi_m(\vec{\gs}_m+(\mathbb{S}-\mathrm{I})\cdot\vec{\gs}_{m
    -n}, g)\,,
\ee
or at the level of individual transseries parameters we find
\be
    \mS_n: \gs_{k,m} \mapsto \gs_{k,m}+\sum_{j=1}^{k-1} \gs_{j,m-n} \fS_{j \to k} \, .
\ee
Having derived the actions of the Stokes automorphisms $\mathfrak{S}_n$, we can now consistently resum the (generic) large order transseries along any of the contours shown in figure \ref{fig:contourcylinder} on the right, running from $t=0$ to $t=+\infty$. Changing the contour prescription can then straightforwardly be compensated for by applying the appropriate automorphism $\mathfrak{S}_n$ to the transseries. For example, as a consequence of \eqref{StokesAutoLOR} we know that the following two resummations are equal:
\be
\cS_{0^+}\left(\sum_{k\geq1} \tilde{\mathsf{S}}_{0\to k}\hat \varphi^{(k)} \re^{-g \log(k)} \right) = \cS_{0^-} \left(\sum_{k\geq1} \mathsf{S}_{0\to k}\hat \varphi^{(k)} \re^{-g \log(k)} \right)\,.
\label{eq:ResummationLORtrans00}
\ee
We can extend this to any of the contours shown on the right of figure \ref{fig:contourcylinder} using our automorphisms $\mathfrak{S}_n$. 

Recall that the original motivation for resumming large order transseries sectors was to decode the higher instanton fluctuations $\varphi^{(k)}_h$ from the large order behaviour of the perturbative coefficients $\varphi^{(0)}_g$. In general we only know those coefficients for nonnegative integers $g$. Within that context, the factor $\re^{-2\pi \ri n g}$ therefore reduces to the identity and all Stokes automorphisms $\mathfrak{S}_n$ are equal to $\mathfrak{S}_0$. Thus, for these purposes, the discussion above simplifies greatly and we can simplify the generic large order transseries \eqref{eq:LORtransseries2} to 
\be
\Phi(\vec{\gs}_0,g)
= \sum_{k=1}^\infty \gs_{k,0}\ \hat \varphi^{(k)}(g) \re^{-g\log(k)}\,.
\ee

\begin{exmp}
\label{ex:discLORtrans}
\begin{leftbar}
We can test the relation \eqref{eq:discLORtransseries} between different discontinuities numerically for the $k=1$ sector of the quartic free energy. As already explained in example \ref{ex:AdlerCylinder}, we do not have a closed form expression for the Borel transform $\cB[\hat{F}^{(1)}](t)$, but we can numerically approximate this function using Padé approximants. In particular, the singularities of the Padé approximant encode the singularities of $\cB[\hat{F}^{(1)}](t)$. To test \eqref{eq:discLORtransseries}, we replaced $\cB[\hat{F}^{(1)}](t)$ by its diaganal Padé approximant of order $N$, $\cB\cP^{(N)}[\hat{F}^{(1)}](t)$, and performed the integration required to compute the discontinuity numerically.

Since this computation does not require the perturbative coefficients, we can test it for any $g$, not necessarily integer. In particular we have tested the relation between the discontinuities at $n=0$ and $n=-1$ for diverse values of $g$. For example, taking $g=5/3$ yields
\begin{align}
\disc_0\hat F^{(1)}(g)
&\approx -0.20236122\ldots+0.00000000\ldots\ri\,,\nn\\
\re^{-2\pi\ri g}\, \disc_{-1}\hat F^{(1)}(g)
&\approx -0.20236165\ldots+0.00000496\ldots\ri\,,\label{eq:testDiscLORtransseries}
\end{align}
which nicely matches to six decimal places. To obtain \eqref{eq:testDiscLORtransseries}, we took the diagonal Padé approximant of order $N=800$. This high order is necessary for producing enough singularities near $\cA^{(1)}_{k, -1}$ to compute $ \disc_{-1}\hat F^{(1)}(g)$ with sufficient accuracy. Including even more terms should increase the accuracy of the match even further.
\end{leftbar}
\end{exmp}

In the next subsection we will see that `horizontal' resummations like \eqref{eq:ResummationLORtrans00}  naturally appear in the derivation of what we will call \textit{exact large order relations}.

\subsection{Exact large order relations}
\label{sec:ELOR}
In the previous subsection we learned that the large order transseries inherits the resurgence structure -- expressed in terms of the Stokes matrix $\bS$ --  from the original transseries, despite its distinct singularity structure. We derived the Stokes automorphism $\mS_n$ for several `horizontal' Stokes transitions that allow us to relate the different resummation prescriptions shown in figure \ref{fig:contourcylinder} on the right. We arrived at these results by starting with the conventional large order relation \eqref{eq:LORinGammas1} that we derived at the end of section \ref{sec:classicallargeorder}. Many properties of the large order transseries that we have discussed can also be derived in a more direct way, as we demonstrate in this section.

Let us therefore take a step back and consider the large order relation in its integral form \eqref{eq:LORinGammas}, which we repeat here for convenience and where we have again discarded the contribution from infinity:
\be
    \frac{\varphi_g^{(0)}}{g!} = -\sum_{k\geq1} \mathsf{S}_{0\to k}\int_0^{\infty -\ri \eps} \frac{\df t}{2\pi \ri} \, \frac{\cB [\varphi^{(k)}](t)}{ (t+kA)^{g+1} } \, .
    \label{eq:coeffexpandint}
\ee
In section \ref{sec:classicallargeorder} we asymptotically expanded the right hand side of this expression by expanding the Borel transform $\cB[\varphi^{(k)}](t)$ as a power series in $t$ and performing the integral term by term. Let us now present a different approach: we first introduce the variable $u = \log(t+kA)$ which allows us to rewrite the expression as
\be
    \frac{\varphi_g^{(0)}}{g!} = -\sum_{k\geq1} \mathsf{S}_{0\to k}\int_{\log(kA)}^{\infty -\ri \eps} \frac{\df u}{2\pi \ri} \, \cB [\varphi^{(k)}](\re^u-kA)\, \re^{-g u} \, .
    \label{eq:ELOR0}
\ee 
Subsequently, we shift the integration variable to $v=u - \log(kA)$ and obtain
\be
    \frac{2\pi \ri \,\varphi_g^{(0)} A^g}{\Gamma(g)} = -\sum_{k\geq1} \mathsf{S}_{0\to k} \re^{-g \log(k)} \, \left(g\int_0^{\infty - \ri \eps} \df v \, \cB[\varphi^{(k)}](kA(\re^v-1))\, \re^{-g v }\right).
    \label{eq:ELOR1}
\ee
On the right hand side we recover the Borel transform \eqref{eq:StirlingTrafoBorel1} of the rescaled Stirling transforms $\hat\varphi^{(k)}$ that we studied before. This equation expresses the perturbative coefficient $\varphi^{(0)}_g$ as the resummation of the large order transseries (\ref{eq:LORtransseries1}) along the contour below the positive real axis. Whenever the sum over $k$ on the right hand side converges, which is generally the case in examples, \eqref{eq:ELOR1} is an exact relation that is also valid for finite $g$. We will therefore call this expression the {\em exact large order relation} from now on. 

The brief computation above reproduces much of the hard work of this section: it allows us to rederive the resurgence relations \eqref{eq:BorelResurgenceLOR} and singularity structure \eqref{eq:LORdefA} and subsequently the automorphisms $\mS$ that we found in the previous section. Most importantly, it assigns a distinct contour $[0^-]$ to our large order transseries with residues $\fS_{0\to k}$ along which its resummation computes the correct value for the perturbative coefficients $\varphi^{(0)}_g$.

The right hand side of the exact large order relation constitutes a function of $g$ that we can expand in a large $g$ limit. Equation~\eqref{eq:ELOR1} tells us that one possible transseries expansion is the large order transseries~\eqref{eq:LORtransseries1}, which after resummation along the $[0^-]$ contour gives back the same function. If we repeat the computation above, but with Hankel contours running above the real positive $t$ axis in \eqref{eq:coeffexpandint} -- see again the bottom graphs in figure \ref{fig:TwoHankels} -- then we obtain a similar expression: 
\be
    \frac{2\pi \ri \,\varphi_g^{(0)} A^g}{\Gamma(g)} = -\sum_{k=1}^\infty \tilde{\mathsf{S}}_{0\to k} \re^{-g \log(k)} \, \left(g\int_0^{\infty+\ri \eps} \df v \, \cB[\varphi^{(k)}](kA(\re^v-1))\, \re^{-g v }\right)\, .
    \label{eq:ELOR2}
\ee
This is simply the $[0^+]$-resummation of the large order transseries with the alternative Borel residues $\tilde \fS_{0\to k}$. The resulting function is of course equal to \eqref{eq:ELOR1}:
\bea
    \frac{2\pi \ri \,\varphi_g^{(0)} A^g}{\Gamma(g)} & = & \cS_{0^+}\left(-\sum_{k=1}^\infty \tilde{\mathsf{S}}_{0\to k} \re^{-g \log(k)} \hat \varphi^{(k)}(g)\right) \ret
    & = & \cS_{0^-} \left(-\sum_{k=1}^\infty \mathsf{S}_{0\to k} \re^{-g \log(k)} \hat \varphi^{(k)}(g)\right)\,.
    \label{eq:ELOR3}
\eea
The equality of both resummations is consistent with the Stokes automorphism \eqref{StokesAutoLOR} that we derived in the previous subsection, and can be seen as an alternative derivation of that equation. More generally, by considering \eqref{eq:ELOR0} and substituting $v=u-\log(kA)+ 2\pi \ri n$ where $n$ is any integer, we can see that
\bea
     \frac{2\pi \ri \varphi^{(0)}A^g}{\Gamma(g)} &=& \re^{-2\pi \ri n g} \, \cS_{n^+}\left(-\sum_{k=1}^\infty \tilde{\mathsf{S}}_{0\to k} \re^{-g \log(k)} \hat \varphi^{(k)}(g)\right) \ret
     &= &  \re^{-2\pi \ri n g}\, \cS_{n^-}\left(-\sum_{k=1}^\infty \mathsf{S}_{0\to k} \re^{-g \log(k)} \hat \varphi^{(k)}(g)\right)\,,
     \label{eq:ELOR4}
\eea
which is consistent with \eqref{eq:ResumNLOR}.  In example \ref{ex:ExactResummingFreeEnergy}, we check the exact relation~\eqref{eq:ELOR1}, and in particular~\eqref{eq:ELOR3}, for the quartic free energy.

\begin{figure}
\centering
\subfloat[]{
    \includegraphics[width=.47\textwidth]{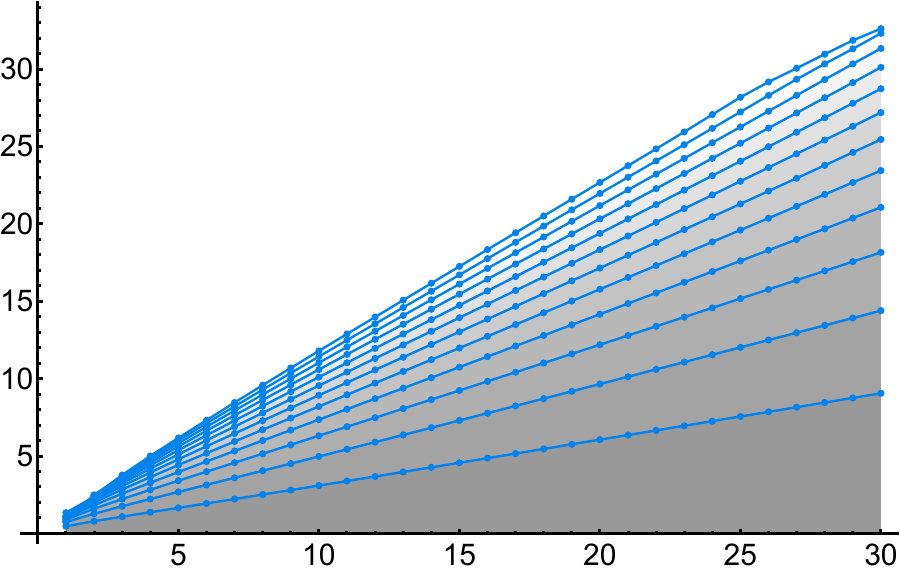}
    \label{fig:resummingFreeEnergyBelow}
}
\subfloat[]{
    \includegraphics[width=.47\textwidth]{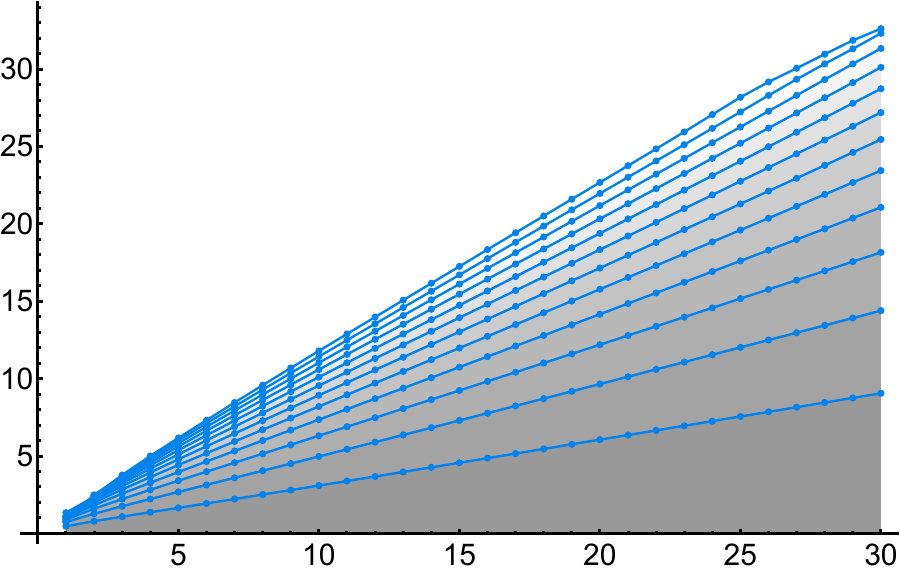}
    \label{fig:resummingFreeEnergyAbove}
}
\caption{In figure~\ref{fig:resummingFreeEnergyBelow} we show the precision -- defined by \eqref{eq:precisionResummingLOR} -- of the resummed large order transseries along the $[0^-]$ contour. We plotted the values for $g$ ranging from 1 to 30 and $n$ ranging from 1 to 12. In figure~\ref{fig:resummingFreeEnergyBelow} we used an integration contour below the real line as in \eqref{eq:resummingLORFree}. In figure~\ref{fig:resummingFreeEnergyAbove} we show a similar plot where we used an integration path along the $[0^+]$ contour instead.}
\label{fig:resummingFreeEnergy}
\end{figure}

\begin{exmp}
\label{ex:ExactResummingFreeEnergy}
\begin{leftbar}
We consider again the quartic free energy. Recall example~\ref{ex:ratioTests}, where we tested the large order transseries \eqref{eq:LORtransseries1} for $k=1$ and in the large $g$-limit by applying ratio tests on
\begin{equation}
r_g 
\equiv \frac{2\pi\ri A^g }{\Gamma(g)} F_g^{(0)}\,.
\end{equation}
We can now also perform tests for $k>1$ by resumming the asymptotic large order relation, i.e. we can test the exact large order relation~\eqref{eq:ELOR1}. 
Similar to example~\ref{ex:discLORtrans}, we do not have an exact Borel transform $\cB[\hat{F}^{(k)}](t)$ at our disposal, so we replace this function by its (diagonal) Padé approximant in the integration and define
\begin{equation}\label{eq:resummingLORFree}
r_g^{(k)} 
\equiv S_1^k\re^{-g\log(k)}\int_0^{\infty-\ri\eps}
dv\,\cB\cP^{[N]}[\hat{F}^{(k)}](v) e^{-g v}\,.
\end{equation}
A numerical evaluation of the integral yields, for example for $g=20$:
\begin{align}
r_{20} 
&\approx \phantom{-0.0000000000000000000+}1.386394325206770660\,\ri\,,\nn\\
r_{20}-r_{20}^{(1)}
&\approx \phantom{-}0.000000879112757638+0.000000000111827370 \,\ri\,,\nn\\
r_{20}-(r_{20}^{(1)}+r_{20}^{(2)})
&\approx \phantom{-}0.000000000000643417-0.000000000223651298 \,\ri\,,\nn\\
r_{20}-(r_{20}^{(1)}+r_{20}^{(2)}+r_{20}^{(3)})
&\approx -0.000000000000643249-0.000000000000010324 \,\ri\,.
\end{align}
To obtain these numbers, we used the $N=100$ diagonal Padé approximant. We observe that $r_{20}^{(1)}$ already matches the exact value $r_{20}$ to $7$ decimal places and adding further $r_{20}^{(k)}$ improves the approximation even further. 

In \cite{Aniceto:2011nu}, a nice way to visualize approximations with an increasing number of nonperturbative sectors was introduced. In figure~\ref{fig:resummingFreeEnergyBelow}, we compare $r_g$ with the resummed large order relation up to twelve nonperturbative sectors in this way. The vertical axis shows the precision, defined by
\begin{equation}\label{eq:precisionResummingLOR}
\log_{10}\left|\frac{r_g}{r_g-\sum_{k=1}^n r_g^{(k)}}\right|\,,
\end{equation}
for $g$ ranging from 1 to 30 and $n$ ranging from 1 to 12.
In figure~\ref{fig:resummingFreeEnergyAbove}, we also test~\eqref{eq:ELOR2} by changing the integration path in \eqref{eq:resummingLORFree} to the $[0^+]$ contour and using the Borel residues \eqref{eq:ResidueQFE2} -- i.e.\ we replace $S_1^k$ by $-(-S_1)^k$. It is important to realize that we must do {\em both} replacements: the large order relation only gives exact results when we use the Borel residues that correspond to the chosen integration contour. We observe that once we do this, both resummations of the large order transseries match the exact values $r_g$ with an extraordinarily small error.
\end{leftbar}
\end{exmp}

In section \ref{sec:classicallargeorder} we discussed the derivation of the classical `asymptotic' large order relation which expresses the large $g$ behaviour of the perturbative coefficients $\varphi_g^{(0)}$ in terms of the large order transseries shown in \eqref{eq:LORtransseries1}. In the present section we learned that we can also express those same perturbative coefficients as the resummation of the large order transseries, without invoking any large $g$ limit. It is therefore natural to wonder if the equation \eqref{eq:ELOR1} is an \textit{exact} large order relation, in the sense that its right hand side computes the perturbative coefficients $\varphi^{(0)}_g$ also at \textit{finite} $g$ to arbitrary precision. Figures~\ref{fig:resummingFreeEnergyBelow} and~\ref{fig:resummingFreeEnergyAbove} support this idea, as both resummations describe the perturbative coefficients with extraordinary precision. Let us discuss this possibility of exactness somewhat further.

One possible objection that one might raise against this assertion is that a finite number of perturbative coefficients $\varphi^{(0)}_g$ can be changed `by hand' by adding a polynomial to $\varphi^{(0)}(z)$, without altering the instanton coefficients that determine the resummation on the right hand side of \eqref{eq:ELOR1}. Therefore, how can that equation be used to exactly compute all $\varphi^{(0)}_g$? As we will explain now, this apparent contradiction is resolved by the contribution at infinity to the Laplace transform that appeared in \eqref{eq:deformLOR} but that we have ignored in what followed. 

Imagine that indeed we wish to alter a select number of perturbative coefficients by adding a polynomial to the original series $\varphi^{(0)}(z)$ as follows:
\be\label{eq:AddPolynomial}
    \tilde \varphi^{(0)}(z) = \varphi^{(0)}(z)+\sum_{g=0}^{G} a_g z^{-g}\,,
\ee
thereby altering the values of $\varphi_g^{(0)}$ with $g \leq G$ to $ \tilde \varphi_g^{(0)}  =\varphi_g^{(0)}+a_g $. Clearly, the large order behaviour of the perturbative coefficients $ \tilde \varphi_g^{(0)}$ is unaffected by such a modification and hence the instanton coefficients $\tilde \varphi^{(k)}_h= \varphi^{(k)}_h$ also remain unchanged. However, the exact large order relation for $\tilde \varphi^{(0)}_g$ at finite $g\leq G$ must differ from the original by $a_g$. To see how this happens, we first note that the Borel transform of our new series $\tilde \varphi^{(0)}$ reads
\be
\cB[\tilde \varphi^{(0)}](t) = \cB[ \varphi^{(0)}](t)+\sum_{h=0}^{G} \frac{a_h}{h!} t^h\,.
\ee
The first term on the right hand side, when inserted into \eqref{eq:deformLOR}, yields the familiar exact large order relation of $\varphi^{(0)}_g$ when we follow the contour deformation as explained in section \ref{sec:classicallargeorder}. The second term on the right hand side, however, leads to poles at $t=0$ and $t=\infty$. If we deform the contour away from zero to $t=\infty$ we obtain
\be
\frac{\tilde \varphi_g^{(0)}}{g!} = \left(\text{exact large order relation of }\varphi^{(0)}_g\right) +  \sum_{h=0}^{G} \frac{a_h}{h!} \frac{1}{2\pi \ri}\oint_{t=\infty}  t^{h-g-1} \, \df t\,. 
\ee
An elementary residue computation shows that the latter term is simply $a_g/g!$, as required. We can even extend the argument to the case that $G=\infty$, as long as the power series that is then added in~\eqref{eq:AddPolynomial} has a finite radius of convergence: the large order behaviour is unaltered and therefore the instanton coefficients of the large order relation will remain unchanged. The Borel transform of such a series with a nonzero radius of convergence will moreover be analytic in the whole complex plane and hence, as with the polynomial, a residue computation at infinity will again simply add the coefficient $a_g/g!$ to the exact large order relation. In the next section, we will discuss a contribution at infinity of this kind in some more detail for the quartic partition function.

A final interesting aspect of the exact large order relation is the fact that, since the right hand side is now a function of $g$, we can carry out the resummation even when $g$ is not a positive integer. In fact, we can resum the large order transseries for arbitrary complex values of~$g$. If we regard the perturbative coefficients as a prescription to assign values $\varphi^{(0)}_g$ to positive integers~$g$, one can wonder whether the resummation of the large order transseries for arbitrary $g$ provides a natural extension of this map $\varphi^{(0)}_g: \mathbb{N} \to \mathbb{C}$ to the complex domain. We will discuss this question for a specific example in the next section.

\section{A worked out example}
\label{sec:ELORforPartitionFunction}
In the previous section we derived the \textit{exact} large order relation~\eqref{eq:ELOR1} for the perturbative coefficients of an asymptotic transseries. Although our derivation of this relation is only valid for integer values of $g$ -- for example because in the Cauchy integral representation (\ref{eq:firststepLOR}) one can only extract residues when $g$ is integer -- nothing prevents us from studying the resulting equation for generic real or even complex values of $g$. In this section, we want to find out to what extent the large order relation is truly exact. We do so by studying the example of the \textit{quartic partition function} -- closely related to the quartic free energy that featured in several previous examples -- for which the perturbative coefficients $\varphi_g^{(0)}$ have a closed form expression which admits a natural extension to complex~$g$. 

In the introduction we stated that one can often think of a transseries as the semiclassical expansion of a path integral around all of its saddle points. To exemplify this, in section~\ref{sec:quarticIntegral} we discuss a well-known toy model called the quartic partition function, which is a zero-dimensional version of the textbook example of $\phi^4$ theory, as well as its logarithm -- the quartic free energy. In section~\ref{sec:LORforPartitionFunction} we verify that the exact large order relation for the quartic paritition function indeed computes the perturbative coefficients, as well as their natural extension to complex values of the index $g$, with arbitrary numerical precision. In the extension to the complex domain, we will see that also for the large order transseries, Stokes phenomenon plays a crucial role. We discuss the interpretation and significance of these results in section~\ref{sec:ELORdiscussion}.

\subsection{Quartic integral and free energy}
\label{sec:quarticIntegral}
So far in this work, we have used the quartic free energy in various examples to illustrate some of the concepts and perform numerical checks on the formulas we derived. For the purpose of the storyline, we kept the examples small and restricted ourselves to the necessary details, postponing a more detailed introduction of the model to the present section. Of course, the quartic model is a favorite toy model in many resurgence studies, so the interested reader can find many further applications of this model elsewhere, notably in \cite{Aniceto_2019}.

Let us start by considering the quartic partition function
\begin{equation}
\label{eq:quarticIntegral}
Z(\hbar)=\frac{1}{2\pi}\int_\Gamma dw\,e^{-\frac1\hbar V(w)}\,,
\end{equation}
where $V(w)$ is a quartic potential defined by
\begin{equation}
\qquad
V(w) = \frac12 w^2-\frac{\lambda}{24} w^4\,,
\end{equation}
and $\Gam$ is an infinite contour in the complex $w$-plane chosen such that the integral converges. The parameters $\hbar$ and $\gl$ can be arbitrary nonzero complex numbers. One can think of \eqref{eq:quarticIntegral} as a zero-dimensional path integral, hence the name `partition function' for $Z(\hbar)$. 

By defining $x\equiv\lambda\hbar$ and changing the variable $w\to\sqrt{\hbar} \, w$, the partition function \eqref{eq:quarticIntegral} can be written as\footnote{For convenience, we change the overall normalisation by a factor $\sqrt{\hbar}/(2\pi)$; see~\cite{Aniceto_2019} for more details.}
\begin{equation}
Z(x) = \frac{1}{\sqrt{2\pi}}\int_{\tilde \Gamma} dw\, e^{-\frac12 w^2 + \frac{x}{24}w^4}\,,
\label{eq:QuarticDef}
\end{equation}
where $\tilde \Gamma$ is the curve obtained from $\Gam$ by the above rescaling. The partition function can be shown to satisfy the second-order linear differential equation 
\begin{equation}
16 x^2 Z''(x)+(32x-24)Z'(x)+3Z(x) = 0\,,
\label{eq:QuarticDiff}
\end{equation}
which is essentially a Ward-Takahashi identity for the zero-dimensional path integral. In the next subsection, we will consider the transseries solution to this differential equation and test the exact large order relation \eqref{eq:ELOR1} for this transseries. 

Before we do so, let us introduce the quartic {\em free energy} $F=\log Z$, which is the model we discussed in the examples \ref{ex:StokesAutoFreeEnergy} and \ref{ex:ratioTests} -- \ref{ex:ExactResummingFreeEnergy}. Using~\eqref{eq:QuarticDiff}, it is straightforward to show that $F$ satisfies the following nonlinear second order differential equation: 
\begin{equation}
16x^2 F''(x)+16x^2\left(F'(x)\right)^2+(32x-24)F'(x) = 0\,.
\label{eq:FreeDiff}
\end{equation}
Being a \textit{second} order ODE, the solution to \eqref{eq:FreeDiff} has two integration constants and is therefore described by a \textit{two}-parameter transseries which turns out to have the form
\begin{equation}
\label{eq:transseriesFreeEnergy}
F(x,\sigma,\rho)
= \sum_{n=0}^\infty \sigma^n\, e^{-\frac{nA}{x}} \sum_{g=0}^\infty F^{(n)}_g x^g +\rho\,,
\end{equation}
with instanton action $A=3/2$ and $\gs$ and $\rho$ the transseries parameters. This expression was used in example~\ref{ex:StokesAutoFreeEnergy}. As \eqref{eq:FreeDiff} is a non-linear ODE, one obtains infinitely many instanton sectors $F^{(n)}$ with evenly spaced actions $n A$, leading to the pattern of Borel plane singularities of figure \ref{fig:residueconvention}. By plugging the transseries \eqref{eq:transseriesFreeEnergy} into \eqref{eq:FreeDiff}, one can obtain recursive relations to determine the instanton coefficients $F_g^{(n)}$ that can be found in appendix B of~\cite{Aniceto_2019}.\footnote{In~\cite{Aniceto_2019}, the instanton $F_g^{(n)}$ are labeled as $F_g^{(n,0)}$. Furthermore, because the integration constant $\rho$ is simply an additive constant, the two-parameter transseries in \eqref{eq:transseriesFreeEnergy} is dubbed a `1.5-parameter' transseries there.}

\subsection{Test of the exact large order relations}
\label{sec:LORforPartitionFunction}
We want to test the exact large order relation~\eqref{eq:ELOR1} for the quartic partition function. We therefore need the solution to the second-order linear ODE (\ref{eq:QuarticDiff}), which takes the form
\begin{equation}
    \cZ(\gs_0, \gs_1, x) = \gs_0 Z^{(0)}(x)+\gs_1\,\re^{-\frac{A}{x}} Z^{(1)}(x)\,,
\end{equation}
with the same instanton action $A = 3/2$ as for the quartic free energy, but now only two asymptotic series:
\begin{align}
Z^{(0)}(x) 
=&\sum_{n=0}^\infty Z_g^{(0)}x^n\,,
\qquad \qquad 
Z^{(0)}_g 
= \left(\frac{2}{3}\right)^g \frac{(4g)!}{2^{6g}\,(2g)! \, g!}\,, \ret
Z^{(1)}(x) 
= &\sum_{n=0}^\infty Z_g^{(1)}x^n\,,
\qquad \qquad 
Z^{(1)}_g 
= -\frac{\ri}{\sqrt{2}}\left(-\frac{2}{3}\right)^g \frac{(4g)!}{2^{6g}\,(2g)! \, g!} \,.
\label{eq:QuarticCoef}
\end{align}
The corresponding $0$-Borel transforms of the two asymptotic sectors can be expressed in closed form in terms of hypergeometric functions and are given by
\begin{align}
\cB [Z^{(0)}](t) 
=&\ {}_2 F_1\left(\frac{1}{4}, \frac{3}{4}, 1; \frac{2t}{3}\right)\,,\ret
\cB [Z^{(1)}](t) 
=& -\frac{\ri}{\sqrt{2}} {}_2 F_1\left(\frac{1}{4}, \frac{3}{4}, 1; -\frac{2t}{3}\right)\,.
\end{align}
With the expressions for the partition function at our disposal, we can now plug these results into the exact large order relation \eqref{eq:ELOR1}.
The exact large order relation for the coefficients $Z_g^{(0)}$ consists of only a single term on the right hand side:
\begin{equation}
Z^{(0)}_g =  -\frac{ S_1}{2\pi \ri}\ \Gamma(g+1)   A^{-g}\  
\int_0^{\re^{-\ri \gt}\infty}\df t\, \cB[Z^{(1)}] \left(\frac{3}{2}(\re^t-1)\right) \re^{-g t} \,,
\label{eq:QuarticELOR}
\end{equation}
where $\gt = \arg(g)$ and the Stokes constant reads $S_1 = -2$. Let us explain the integration path in \eqref{eq:QuarticELOR}. The extension of our exact large order relation to generic real and positive $g$ is straightforward, but in the case of complex $g$ we need to be careful: when $\Re(g)$ is negative, any Laplace integral going from $0$ to $+\infty$ along the real $t$-axis will no longer converge. Hence we need to adjust the angle along which we integrate as we move $g$ through the complex plane. The correct way to do this is to set the angle of integration to $-\gt$. This will moreover ensure that the Laplace integral converges as rapidly as possible, which is desirable from a numerical computational point of view.

\begin{figure}
\centering
\subfloat[$\arg(g) = 0$]{
    \includegraphics[width=.47\textwidth]{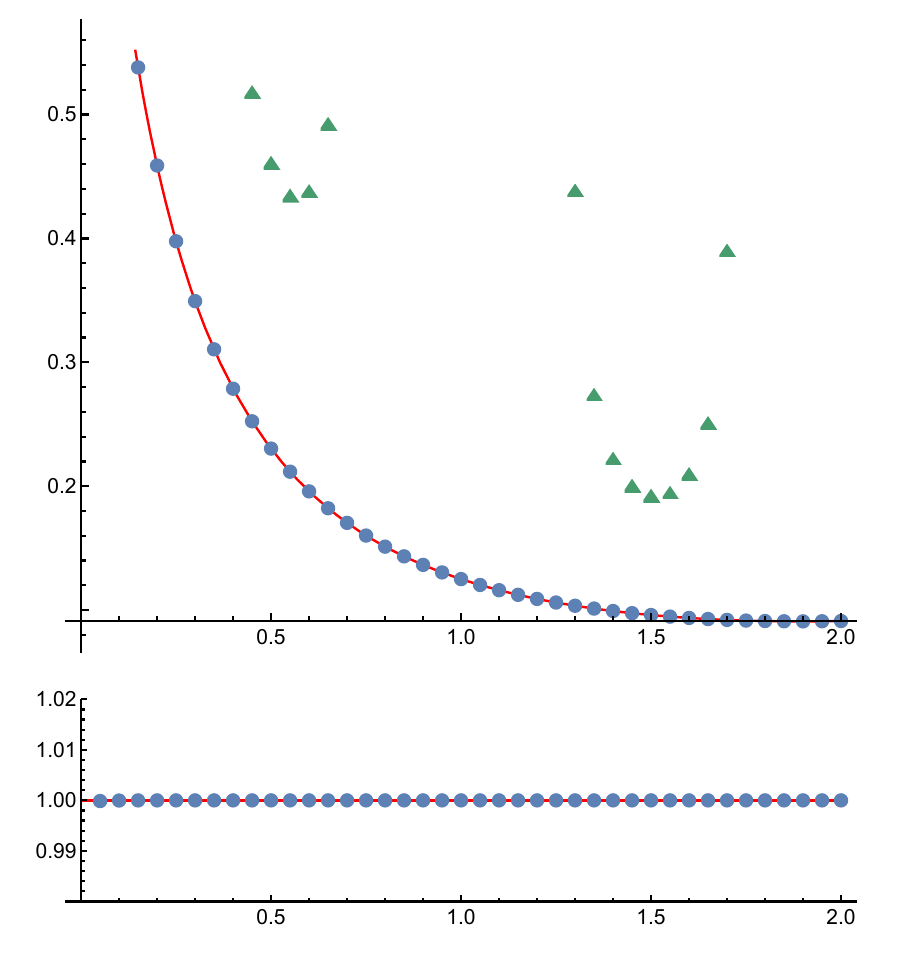}
    \label{fig:ELOR_1_partition_function}
}
\subfloat[$\arg(g) = 0.45\pi$]{
    \includegraphics[width=.47\textwidth]{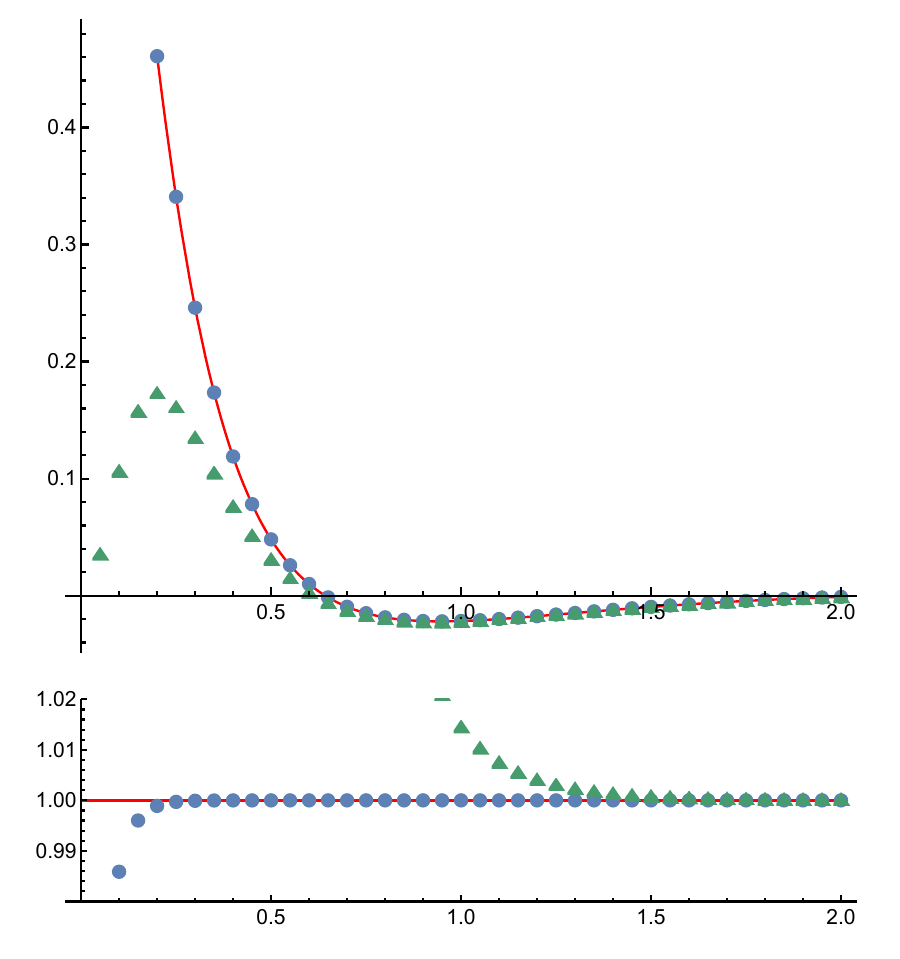}
    \label{fig:ELOR_2_partition_function}
}\\
\subfloat[$\arg(g) = 0.55\pi$]{
    \includegraphics[width=.47\textwidth]{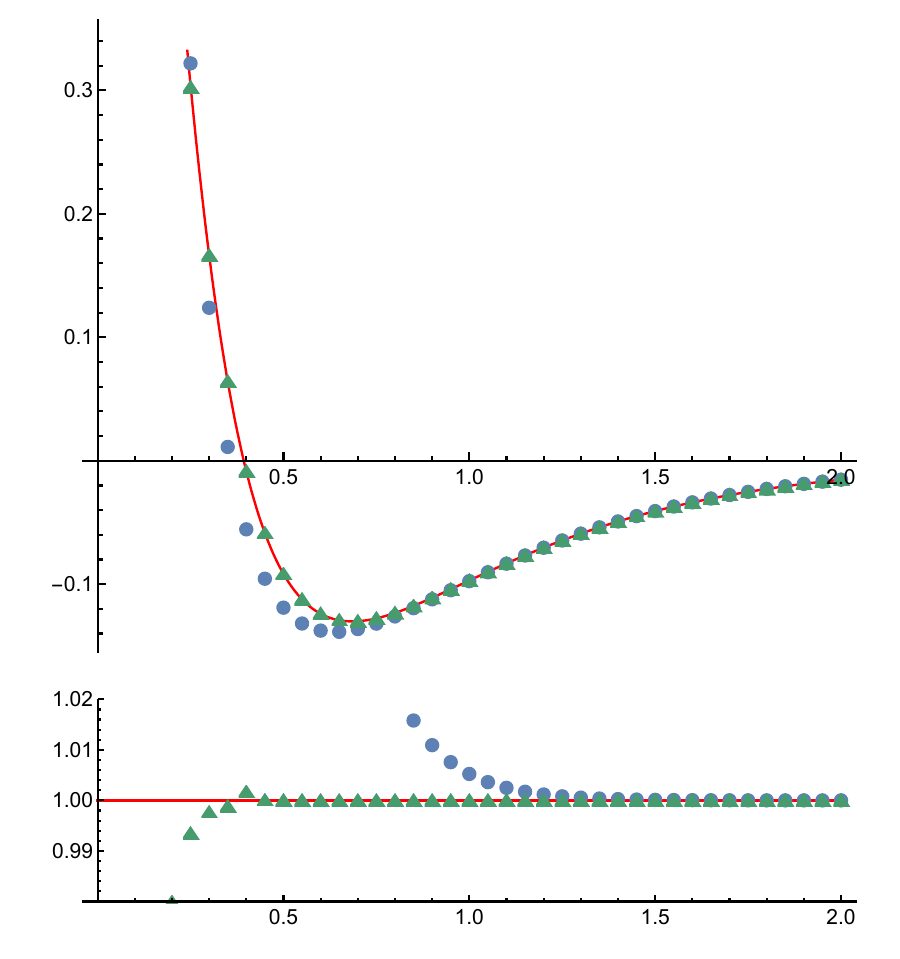}
    \label{fig:ELOR_3_partition_function}
}
\subfloat[$\arg(g) = \pi$]{
    \includegraphics[width=.47\textwidth]{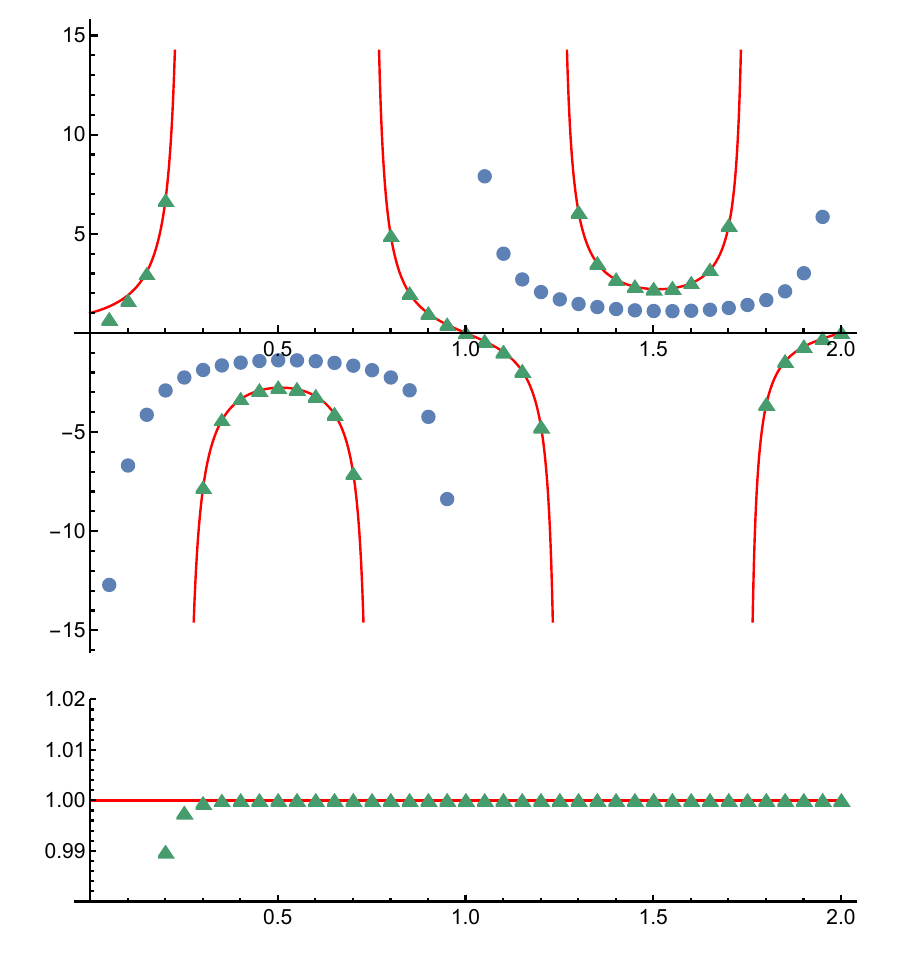}
    \label{fig:ELOR_4_partition_function}
}
\caption{Results of the numerical integration of the exact large order relation for the quartic partition function for four different values of $\arg(g)$ and with $|g|$ (horizontal axis) ranging from 0 to~2.
In the upper graphs in each panel, we show in red (solid line) the real part of the \textit{exact} value \eqref{eq:QuarticCoefAC} of $Z^{(0)}_g$. 
The round blue markers display the numerical results for the exact large order relation \eqref{eq:QuarticELOR}, whereas the green triangle markers show the results for \eqref{eq:QuarticELOR2}, with Stokes phenomenon taken into account. The Stokes phenomenon, which occurs at $\arg(g)=\pi/2$, explains why the blue markers match the red line in figures~\ref{fig:ELOR_1_partition_function} and~\ref{fig:ELOR_2_partition_function}, and the green markers do so in~\ref{fig:ELOR_3_partition_function} and~\ref{fig:ELOR_4_partition_function}.
In the lower graphs in each panel, we show the ratio of the exact and computed values. We achieve a good accuracy for $|g|>0.5$, whereas for smaller values of $|g|$, the numerical accuracy decreases due to the exponential factor in the Laplace integral which becomes large in this regime.
In figure~\ref{fig:ELOR_4_partition_function}, we shifted $|g|$ by $1/1000$ in the numerical integrations to avoid computing ratios of infinities. The lower graph shows that even close to the singularities, the exact large order relation works to great accuracy.}
\label{fig:Elor_partition_function}
\end{figure}
If we regard the perturbative coefficients $Z^{(0)}_g$ in (\ref{eq:QuarticCoef}) as obtained from a function whose domain consists of integers $g\in \mathbb{N}$, then this function admits a natural extension to the domain of complex numbers:
\be
    Z^{(0)}_g = \left(\frac{2}{3}\right)^g\frac{\Gamma(4g+1)}{2^{6g}\,\Gamma(2g+1)\Gamma(g+1)}\,,
    \label{eq:QuarticCoefAC}
\ee
i.e.\ we replace the factorials by gamma functions.\footnote{Strictly speaking we also need to define $c^g$ for constant $c$ as $e^{g \log c}$ and pick a branch for the logarithm, but we can ignore this subtlety in what follows and simply assume that $c^g$ is always extended in the `natural' way from integer $g$.} The first question that we want to address is whether the exact large order relation can reproduce the exact value of $Z^{(0)}_g$, defined in this way, for real positive $g$. The answer to this question is shown in the upper graph of figure \ref{fig:ELOR_1_partition_function} for the real part of $Z^{(0)}_g$. (One obtains similar plots for its imaginary part.) The solid red line plots the \textit{exact value} (\ref{eq:QuarticCoefAC}), and the round blue markers are computed using the right hand side of (\ref{eq:QuarticELOR}). The triangular green markers in this plot will be explained shortly. The lower graph depicts the ratio between the exact and computed values. Clearly, the two expressions match nicely also for non-integer $g$. Notice that there are no singularities on the integration path -- i.e. on the real positive line -- of \eqref{eq:QuarticELOR}, so we do not need to deform the contour here.

We can also consider \textit{negative real} values of $g$. For this, we need to rotate the phase of $g$ from $0$ to $\pi$ and rotate the direction of the integration contour in (\ref{eq:QuarticELOR}) to $-\infty$ in the opposite direction, as we explained before. In figure \ref{fig:ELOR_4_partition_function} we show the result. At first sight our conjecture of exactness of (\ref{eq:QuarticELOR}) seems to fail, as the blue markers clearly fail to accurately describe the analytic expression for $Z^{(0)}_g$. However, there is a simple reason for this: the Borel transform $\cB [Z^{(1)}]\left(\frac{3}{2}(\re^t-1)\right)$ has numerous singlarities along the imaginary axis, and these induce a Stokes phenomenon when the integration contour crosses this axis. Hence, when we rotate the contour of integration to the negative axis in the Borel plane, we pick up nonperturbative corrections that we need to include. Taking these nonperturbative contributions into account we find for $\Re(g)$ \textit{negative} that
\be
Z^{(0)}_g =  -\frac{ S_1}{2\pi \ri}\ \Gamma(g+1) A^{-g} \left(1-\frac{2\re^{2\pi \ri g}}{1+\re^{4\pi \ri g}}\right)\  \int_0^{\re^{-\ri \gt}\infty} \cB Z^{(1)}\left(A(\re^t-1)\right) \re^{-g t} \df t\,,
\label{eq:QuarticELOR2}
\ee
The nonperturbative corrections are therefore described by including the additional factor 
\begin{equation}
-\frac{2\re^{2\pi \ri g}}{1+\re^{4\pi \ri g}}\,,    
\end{equation}
as we derive in appendix \ref{sec:AppGamma}. Using the expression \eqref{eq:QuarticELOR2} we obtain the \textit{green} (triangular) markers in the different panels of figure~\ref{fig:Elor_partition_function}. Indeed, these markers describe the exact values of the perturbative coefficients for $\Re(g)<0$, as one can see in figure \ref{fig:ELOR_4_partition_function}.

Now, we can also test the exact large order relation for complex values of $g$. The most interesting values are those for which the integration contour in the Borel plane is near the Stokes line. To clearly see the Stokes phenomenon, we have therefore tested the exact large order relation for $\arg(g) = 0.45\pi$ (figure \ref{fig:ELOR_2_partition_function}) and $\arg(g) = 0.55\pi$ (figure \ref{fig:ELOR_3_partition_function}). Before the Stokes phenomenon takes place -- which happens at $\gt = 0.5\pi$ -- we observe that the blue markers obtained using \eqref{eq:QuarticELOR} correctly describe $Z^{(0)}_g$. Conversely, the green markers \eqref{eq:QuarticELOR2} describe $Z^{(0)}_g$ correctly after the Stokes phenomenon has occurred. Note that in particular the difference between the blue and green markers in figures~\ref{fig:ELOR_2_partition_function} and~\ref{fig:ELOR_3_partition_function} becomes more significant when $g$ is very small, since -- as is always the case for a Stokes transition -- the two differ by a nonperturbative, exponentially small contribution in the large $g$ expansion.

\subsection{Discussion}
\label{sec:ELORdiscussion}
The analysis in this section has shown that for the quartic partition function, the exact large order relation \eqref{eq:QuarticELOR2} describes the natural extension \eqref{eq:QuarticCoefAC} of the perturbative coefficients to the complex $g$-plane to high numerical precision. To make this work, we needed to account for the Stokes phenomenon, showing that this phenomenon plays an equally important role for large order transseries as it does for `ordinary' ones. 

It is important to realize, though, that as the perturbative coefficients $\varphi_g^{(0)}$ are only defined for positive integers $g$, the extension to complex $g$ is in general not unique. The extension of the factorials to gamma functions was the natural choice -- in a sense, we make the same choice as one does when using the Laplace integral (which defines the gamma function) in \eqref{eq:laplacedef} to reintroduce the factorials after having performed a Borel transform. Less natural options are certainly available: one could for example extend the factorials to Hadamard's gamma function, but with this replacement \eqref{eq:QuarticELOR} and \eqref{eq:QuarticCoefAC} would no longer lead to the same result.

One may now wonder whether the exact large order relation for complex $g$ only worked for this specific example, or if it works in general. Let us first emphasize that for more complicated (non-linear) examples one cannot even give a `natural' extension of $\varphi_g^{(0)}$ to complex values of $g$. Furthermore, deriving the integral expression of the exact large order relation by other means that provide a natural smooth $g$-dependence -- as we were able to do in appendix~\ref{sec:AppGamma} for~\eqref{eq:QuarticELOR} -- is not possible in general.  
In fact, we can turn things around and say that the exact large order relation itself defines the most natural extension to the complex $g$-plane of $\varphi_g^{(0)}$. 
Of course, as the example of the partition function shows, if we want to {\em define} an extension to complex $g$ in this way, Stokes pheneomenon must be taken into account carefully. A final subtelty is that we have ignored a potential contribution at infinity in the Cauchy integral that is used to extract $\varphi^{(0)}_g$ from the Borel transform, also when we took $g$ non-integer. Let us now discuss this subtlety for the quartic parition function.

To this end, let us ask under which conditions a contribution at infinity of the form
\be
    \oint_{\infty} \frac{\df t}{2\pi \ri} \frac{\cB[ \varphi^{(0)}](t)}{t^{g+1}} \, ,
    \label{eq:contributioninfty}
\ee
is nonzero. In the explicit example of the quartic partition function, we already found a match between the resummed large order transseries without such a contribution and the exact value of the perturbative coefficients for values of $g$ throughout the complex plane. This seems to suggest that in this example the contribution at infinity vanishes for any $g$, but is this really the case?

Because we have an explicit expression for the Borel transform of the quartic partition function, we can expand it around $t= \infty$ and obtain 
\be
    \cB [Z^{(0)}](t) \simeq t^{-\frac{1}{4}}\, \sum_{n=0}^\infty a_n \left(\frac{1}{\sqrt{t}}\right)^n\, ,
    \label{eq:B0Z0atinfinity}
\ee
where the coefficients $a_n$ depend on a branch choice at infinity of the hypergeometric function. Since the Borel transform goes to zero as $t$ approaches infinity, the integral \eqref{eq:contributioninfty} vanishes for $\Re(g)>0$: consider a circular path $C_R$ of radius $R$, then the integral scales as
\be
     \int_{C_R} \frac{\df t}{2\pi \ri} \frac{\cB[Z^{(0)}](t)}{t^{g+1}} \simeq \cO\left(R^{-g-1/4}\right)\, ,
     \label{eq:scalingbehaviour}
\ee
which for $\Re(g)>0$ vanishes as $R\to \infty$. When we complexify $g$ and rotate into the half-plane where $\Re(g)<0$, the situation is complicated. Formally, the contribution at infinity should diverge and therefore the exact large order relation should break down. However, this is not what we found in this section: remarkably, the exact large order relation seems to hold, even when $\Re(g)<0$, if we account for the Stokes phenomenon of $\cB[Z^{(1)}]\left(\frac{3}{2}(\re^t-1)\right)$. It is noteworthy, however, that the integrand of \eqref{eq:contributioninfty}, when we plug in the expansion \eqref{eq:B0Z0atinfinity}, has a simple pole -- with residue $a_n$ -- at infinity whenever $g = -\frac{1}{4}-\frac{n}{2}$ for a nonnegative integer $n$. These are exactly the values of $g$ for which both sides of the large order relation \eqref{eq:QuarticELOR2} diverge and where a finite contribution $a_n$ would be neglectable anyway -- see also figure \ref{fig:ELOR_4_partition_function}. Formally, this argument is invalidated by the fact that the residue is defined as the coefficient of the simple pole in a Laurent series expansion in $t$, which the above series is not -- it is an expansion in fractional powers of $t$. Yet, the fact that the series has simple poles for the same values of $g$ for which the coefficient $Z^{(0)}_g$ diverges, does not seem like a coincidence. 

It would be interesting to better understand the nature of the contribution at infinity and how in the end it does or does not contribute to the exact large order relation for complex values of $g$, as well as to what extent this holds true for other transseries. In this regard, note that this example is much simpler than generic large order transseries, since there is only a single instanton sector and hence no Stokes transitions $\mS_n$ across horizontal lines on the Borel cylinder to account for. In fact, `unwrapping' the Borel cylinder for a fully nonlinear problem, one would find an infinite number of diagonal Stokes lines in the large order Borel plane, each containing a single primitive instanton action plus all its positive integer multiples. Combining all of the corresponding Stokes automorphisms into a single one that rotates from the positive to the negative real $t$-axis seems to be a very difficult task.

Finally, let us comment on the nature of the Stokes phenomenon, computed in \eqref{eq:StokesPhenB0Z1}, of the Borel transform $\cB[\hat Z^{(1)}](t) \equiv \cB[Z^{(1)}] \left(\frac{3}{2}(\re^t-1)\right)$ in the example of the quartic partition function. Usually, when an asymptotic series undergoes a Stokes phenomenon, it `jumps' by an exponentialy small factor $\re^{-Az}$ multiplied by a \textit{diferent} asymptotic series. On the contrary, here we observe that the series $\hat Z^{(1)}(x)$ jumps by an exponentially small factor multiplied by \textit{itself}. This phenomenon is known as \textit{self-resurgence} and appears also in the context of WKB solutions (see e.g.\ section 5 of \cite{inbook}). Using asymptotic large order relations, we have checked explicitly that the higher order terms of the formal power series $\hat Z^{(1)}(x)$ probe the lower order terms of that exact same series. From the dependence on $e^t-1$, one sees that the singularities at $t= 2\pi \ri n$ for nonzero integers $n$ in the Borel cylinder of $\cB[\hat Z^{(1)}]$ correspond to the origin in the original Borel plane of $\cB[Z^{(1)}]$. This shows that there is actually a nonzero Borel residue at the origin on the other sheets of the multisheeted (original) Borel plane -- i.e.\ if one takes a path around some of the other singularities. This is a particular example of the multivaluedness of the Borel residues that we discussed in section \ref{sec:AmbStokesData}. This particular type of Stokes phenomenon for large order transseries appears in this example because both $Z^{(0)}$ and $Z^{(1)}$ know about each other -- that is: $\fS_{0 \to 1}$ and $\fS_{1 \to 0}$ are both nonzero. In the context of nonlinear ODEs, one always has that the residues $\fS_{k\to 0 }$ vanish, and hence the same effect does not occur in the Borel cylinders of the first sector $\cB[\hat \varphi^{(1)}]$ -- see e.g. figures \ref{fig:BPplotAdler1}, \ref{fig:BPplotAdler2} and \ref{fig:BP_plot_free_energy}. Even in those examples, though, it does occur for the higher sectors $\cB[\hat \varphi^{(k>1)}]$ -- see e.g. figure \ref{fig:BP_plot_free_energy2}, where we do find singularities at $t=2\pi\ri n$ for integer $n\neq 0$.

\section{Conclusion}
\label{sec:conclusion}
Large order relations are a powerful tool for decoding fluctuations in higher nonperturbative sectors, for example around instantons. They have been used extensively in a wide variety of physical models. In this work, originally motivated by the desire to understand the repeating singularity structure found in figure \ref{fig:BPplotAdler}, we have taken a closer look at the generic underlying transseries structure of these large order relations. We have argued that for rather generic transseries --  only assuming the resurgent structure \eqref{eq:BorelResurgenceGeneral} --  we can fully explain this observed singularity structure, as described by \eqref{eq:LORdefA}. We have seen that the large order transseries is constructed out of rescaled Stirling transforms $\hat \varphi^{(k)}$ of the original instanton sectors, and that these formal power series have a Borel cylinder rather than a Borel plane: $t \sim t+2 \pi \ri$. 

We derived the Stokes automorphism $\mS_0$ of the large order transseries across the positive real axis of this Borel cylinder, which turns out to be qualitatively the same as that of the original transseries, expressed in terms of the Stokes matrix $\bS$ in \eqref{eq:StokesMatrix}. After establishing further automorphisms $\mS_n$ across copies of the positive real axis, we obtained a clear and unambiguous resummation procedure for the large order transseries which, after establishing the exact large order relations \eqref{eq:ELOR4}, is expected to compute the perturbative coefficients $\varphi^{(0)}$ {\em exactly} in many cases.
We have checked the `exactness' of this statement for the quartic partition function, where we found that the exact large order relation holds, even for complex values of $g$, and that Stokes phenomenon for the large order transseries plays an important role to make this happen.

From a practical point of view, what we have gained is a clear understanding of the precise contour along which one is supposed to resum the large order transseries sectors with a given set of Borel residues in order to succesfully probe higher order instanton coefficients $\varphi^{(k>1)}_h$. Having established a clear understanding of its Stokes phenomenon, we can now also resum the large order transseries along other contours using the automorphisms $\mS_n$.  We would like to stress that our analysis makes no assumptions about the Stokes data of the underlying transseries: we formulated our results in terms of generic Borel residues rather than specific Stokes constants.

From a more conceptual point of view, it is interesting that the \textit{exact} large order relation does not only compute the perturbative coefficients $\varphi^{(0)}_g$, but also allows us to extend them to complex values of $g$. It is striking that the exact large order relation for the quartic partition function, which we tested thoroughly in section \ref{sec:ELORforPartitionFunction}, seems to work for \textit{all} complex values of $g$ -- if we take the Stokes phenomenon explained in appendix \ref{sec:AppGamma} into account -- and does not seem to be spoiled by the contribution at infinity in the Cauchy integral. Of course, this is only a single example, but it would be interesting to see more examples where this contribution might or might not be relevant.

A first open question that would be interesting to address is therefore whether the example of the quartic partition function is special or whether the exactness that we found holds for large classes of examples, either because contributions from infinity are absent or because we can pinpoint such contributions and correctly take them into account. This question is intimately related to the question whether our techniques (and in particular the Stirling transform) can lead to numerical gain: while from the computations that we have performed the use of the Stirling transform does not seem to lead to much faster numerics for large values of $g$, it may lead to improvements for small values of $g$ -- but to compute at such values, it is crucial that many contributions to the exact large order relation -- including potential contributions from infinity -- are properly taken into account.

The results of this paper can most likely be generalized to broader settings, and it would be interesting to explore some of those. Here, one can think of transseries with multiple Stokes lines, transseries with logarithmic terms (such as resonant transseries), transseries that have additional parameters and display the higher order Stokes phenomenon of \cite{Howls2004}, etc. While we do not expect major qualitative differences in such cases, it would still be good to extend the large order toolbox that we have developed to these settings.

The geometry of the Borel cylinder also begs to be explored further. While the $t \sim t+2 \pi \ri$ symmetry in the Borel plane of the large order transseries is exact when we explore positive integer $g$, in more general cases our expressions involved factors of $e^{-2\pi\ri g}$ for each shift in the imaginary direction. This seems to hint at a structure where the quantities that we compute, rather than functions on the Borel cylinder, are sections of a bundle over this cylinder. Exploring this possibility may lead to an even better, more geometric understanding of large order relations. Another geometric question arises from the fact that the locations of the singularities on the Borel cylinder are the logarithms of the locations of the singularities of the original transseries. This may suggest that we should not only think of sectors $\hat \varphi^{(k)}$ associated to singularities at $\log(k)$, but that one can also think of a sector $\hat \varphi^{(0)}$ that encodes the original perturbative series and that has been shifted to the end of the Borel cylinder at $-\infty$.

\acknowledgments
We thank David Sauzin for useful correspondence and for sharing an unpublished worked out example with us. The research of AvS was supported by the grant OCENW.KLEIN.128, ‘A new approach to nonperturbative physics’, from the Dutch Research Council (NWO).

\appendix

\section{\texorpdfstring{$n$}{n}-Borel transforms}
\label{sec:nBorel}
Assume that we are given a sequence of complex numbers $\varphi = \{ \varphi_g \, | \, g \in \bZ_{\geq 0} \}$. We will think of these coefficients as coefficients of some power series, e.g.\
\be
\qquad \varphi(z) = \sum_{g=0}^\infty \varphi_g z^{-g}\, ,
\ee
but for most purposes in this appendix the exact origin of the sequence $\varphi$ is irrelevant.

We define the {\em $0$-Borel transform} of the sequence $\varphi$ as
\be
 \cB_0[\varphi](t) \equiv \sum_{g=0}^\infty \frac{\varphi_g}{g!} \, t^g\, .
\ee
The $0$-Borel transform is a formal power series in the variable $t$, but as usual, when this power series converges in some domain we use the same notation for the function to which it converges, as well as for its analytic continuation.

The {\em $1$-Borel transform} of $\varphi$ is then defined to be the $t$-derivative of the $0$-Borel transform:
\be
 \cB_1[\varphi](t) \equiv \frac{d}{dt} \cB_0[\varphi](t) = \sum_{g=0}^\infty \frac{\varphi_{g+1}}{g!} \, t^{g}\, .
\ee
More generally, we define the {\em $n$-Borel transform} of $\varphi$ to be
\be
 \cB_n[\varphi](t) \equiv \sum_{g=0}^\infty \frac{\varphi_{g+n}}{g!} \, t^{g}\, ,
 \label{eq:nBoreldef}
\ee
which can be viewed as the $n$th derivative of the $0$-Borel transform of $\varphi$. In fact, for general real (or even complex) $\nu$, one may define
\be
 \cB_\nu[\varphi](t) \equiv \sum_{g=0}^\infty \frac{\varphi_g}{\Gamma(g-\nu+1)} t^{g-\nu}\, ,
\ee
which for $\nu = n$ integer agrees with (\ref{eq:nBoreldef}). It is then generally true that
\be
 \frac{d}{dt} \cB_\nu[\varphi](t) = \cB_{\nu+1}[\varphi](t)\, .
 \label{eq:borelder}
\ee
Once a $\nu$-Borel transform is analytically continued beyond the disk of convergence, it can be mapped back to the \textit{Borel sum} via
\be
    \cS_\gamma \varphi(z) = z^{1-\nu}\int_\gamma \cB_\nu[\varphi](t) \re^{-z t} \df t \,,
    \label{eq:BorelSum1}
\ee
where to ensure convergence, the contour $\gamma$ is a straight path starting at the origin and approaching infinity in the half plane where $\Re(zt)>0$. Note that the left hand side of this equation is independent of $\nu$. This resummation constructs a function of $z$ which we can asymptotically expand again to retrieve the formal power series $\varphi(z)$. We could also rewrite the Borel sum \eqref{eq:BorelSum1} as
\be
 \cS_\gamma \varphi(z)= z^{-\nu}\int_0^\infty \cB_\nu[\varphi]\left(\frac{s}{z}\right) \re^{-s} \df s
 \label{eq:BorelSum2}
\ee
and \textit{fix} the integration contour to run from zero to infinity along the positive real axis. Under such conventions, convergence is ensured and the locations of the Borel singularities become dependent on $z$. We then find that Borel singularities will lie on the positive real $s$-axis for certain values of $z$ that paramatrize a set of lines in the complex $z$-plane that are often also called Stokes lines. In this work however, we use \eqref{eq:BorelSum1} and employ a definition in which Stokes lines are rays in the Borel $t$-plane on which singularities lie.

When we start from a sequence that grows factorially as
\be
 \frac{\varphi_g}{A^{-(g+\beta)} \Gam(g+\beta)} = \mbox{const} + \cO\left(\frac{1}{g}\right) \qquad \mbox{as~} g \to \infty
 \label{eq:betagrowth}
\ee
for some nonzero complex constant, and some real choice of $\beta$, then $\cB_{1-\beta}[\varphi](t)$ has a simple pole at $t=A$. Using (\ref{eq:borelder}), we see that similarly, for integer $k \geq 1$, $\cB_{k-\beta}[\varphi](t)$ has a pole of order $k$ at $t=A$, and $\cB_{-\beta}[\varphi](t)$ has a logarithmic branch point. We use these facts in this paper to assume, without too much loss of generality, that our functions in the Borel plane have only logarithmic branch points as their singularities -- for any given Gevrey-1 asymptotic power series and any given singular point $t=A$, we can always choose an appropriate $n$-Borel transform such that this is the case.

Of course, the above discussion should not be taken to imply that {\em all} sectors in a given transseries have logarithmic branch points at {\em all} singularities in the Borel plane: the Painlevé I equation is an example where the value of $\beta$ in \eqref{eq:betagrowth} differs from sector to sector and has no upper bound -- see \cite{Aniceto:2011nu} -- so that there will always be sectors where poles in the Borel plane do appear. Extending the results of this paper so that they also apply to these cases is completely straightforward: we insist on logarithmic branch points only to keep our formulas concise. Similarly, it is straightforward to derive versions of most formulas appearing in this paper for cases where $\beta$ is non-integer.


\section{Properties of Stirling transforms}
\label{sec:AppBorelStirling}
Stirling numbers of the second kind can be defined as \cite{NIST:DLMF}
\be
    \stir{n}{k} = \frac{1}{k!}\sum_{j=0}^k(-1)^j \binom{k}{j}\left(k-j\right)^n\,.
    \label{eq:StirnumberDef}
\ee
They allow us to express the $0$-Borel transform of the Stirling transform $\tilde{\psi}(g)$ defined in \eqref{eq:Stirdefg} in terms of the $0$-Borel transform of $\psi(z)$:
\be
    \cB_0[\,\tilde \psi\,](t) = \sum_{n=0}^\infty \sum_{k=0}^\infty \sum_{j=0}^k\frac{(-1)^j (k-j)^n a_k }{k! n!} \binom{k}{j} t^n = \cB_0 [\,\psi\,](\re^t-1)\,.
    \label{eq:BorelgtoBorelf}
\ee
The $n$-Borel transform is then obtained by simply differentiating both sides $n$ times with respect to $t$. One finds for example 
\be
    \cB_1 [\,\tilde \psi\,](t) = \re^t \cB_1 [\,\psi\,](\re^t-1)\,,
\ee
which can also be obtained by explicitly computing $\cB_1[\tilde \psi]$ using (\ref{eq:nBoreldef}) and (\ref{eq:StirnumberDef}), just as we showed for the $0$-Borel transform in (\ref{eq:BorelgtoBorelf}). The $0$-Borel transform of the rescaled Stirling transform \eqref{eq:Stirdefgtil} is essentially obtained by repeating the computation (\ref{eq:BorelgtoBorelf}), only with $a_k$ replaced by $a_k A^k$, leading to 
\be
    \cB_0[\,\hat \psi\,](t) = \cB_0 [\,\psi\,]\left(A(\re^t-1)\right)\,,
\ee
which again can be differentiated with respect to $t$ to obtain the corresponding $n$-Borel transforms.

\section{Stokes phenomenon for gamma functions}
\label{sec:AppGamma}
\begin{figure}
    \centering
    \includegraphics[width = 0.8\linewidth]{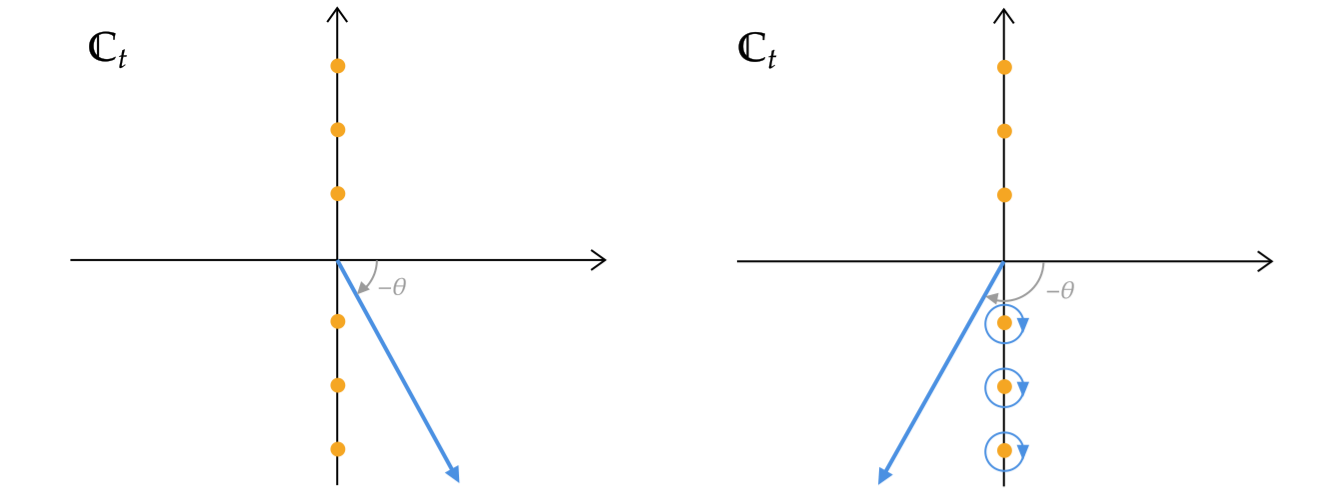}
    \caption{The Stokes phenomenon of $\cF_1$}
    \label{fig:DiscGamma}
\end{figure}

In the example of the quartic partition function, the Borel transform $\cB [Z^{(1)}]\left(\frac{3}{2}(\re^t-1)\right)$ induces a Stokes phenomenon as we cross the imaginary $t$-axis. In order to derive the exact form of this Stokes phenomenon, we start from the exact large order relation as it holds for positive $g$:
\be
    \int_0^\infty \cB [Z^{(1)}]\left(\frac{3}{2}(\re^t-1)\right) \re^{-g t} \df t = -\frac{2\pi \ri}{S_1} \, \frac{\Gamma(4g+1)}{2^{6g}\, \Gamma(2g+1)\Gamma(g+1)^2}\,.
    \label{eq:D1}
\ee
Our goal is then to derive the Stokes phenomenon that the gamma functions on the right hand side of this expression undergo and combine them into a Stokes phenomenon for the whole expression.
To this end, we use Binet's Log Gamma formula,
\be
\log \Gamma (z) = \underbrace{\left(z-\half\right)\log(z) - z +\half \log(2\pi)}_{C(z)} + \int_0^\infty \left(\half - \frac{1}{t}+\frac{1}{\re^t-1}\right)\frac{\re^{-z t}}{t} \df t\,,
\ee
for $\Re(z)>0$. This expression consists of singular and constant terms (as $1/z \to 0$) that we collect in the function $C(z)$, and $1/z$ corrections which are resummed using Borel-Laplace resummation. Let us first shift the argument by one:
\be
\log \Gamma (z+1) = C(z+1) + \int_0^\infty \left(\left(\half - \frac{1}{t}+\frac{1}{\re^t-1}\right)\frac{\re^{-t}}{t}\right) \re^{-zt}\df t\,.
\ee
In the latter term we interpret the integrand as a 1-Borel transform of some formal power series $F(z)$:
\be
    \cB_1[ F](t) =  \left(\half - \frac{1}{t}+\frac{1}{\re^t-1}\right)\frac{\re^{-t}}{t}\,.
\ee
By shifting and rescaling the argument we straightforwardly find that 
\bea
    \log \Gamma (z+1) &=& C(z+1)+\int_0^\infty \cB_1[ F]\left(t\right) \re^{-zt}\df t \,, \ret
    \log \Gamma (2z+1) &=& C(2z+1)+\int_0^\infty \half\cB_1 [F]\left(\frac{t}{2}\right) \re^{-zt}\df t \,, \ret
    \log \Gamma (4z+1) &=& C(4z+1)+\int_0^\infty \frac{1}{4}\cB_1 [F]\left(\frac{t}{4}\right)  \re^{-zt}\df t\,. 
\eea
The large $z$ expansion of the function $C(z)$ has a finite radius of convergence and will therefore not contribute to any Stokes phenomenon that the large $z$ expansion of $\log \Gamma(cz+1)$ undergoes. Let
\be
    \cB_1 [\cF_n](t) = \frac{1}{n}\cB_1 [F]\left(\frac{t}{n}\right)\,,
\ee
then each $\cB_1 [\cF_n](t)$ has simple poles located at $t=2\pi  \ri n m$, with residue $1/(2\pi \ri m)$, for all nonzero integers $m$. When we move the contour of integration across the imaginary $t$-axis -- as depicted in figure \ref{fig:DiscGamma} for $\cF_1$ -- we therefore pick up poles which lead to the discontinuity
\be
    \text{disc}_{-\frac{\pi}{2}} \cF_n = -\log\left(1-\re^{2\pi \ri n g }\right)\,.
\ee
Note that the contour of integration runs along an angle $-\gt$ in the lower half plane in order to keep $\re^{-gt}$ real and exponentially suppressed. The Stokes automorphism on the product of gamma functions then acts as
\be
    \mathfrak{S}_{-\frac{\pi}{2}} \re^{\cF_4-\cF_2-2\cF_1} = \left(1-\frac{2\re^{2\pi\ri g}}{1+\re^{4\pi \ri g}}\right)\re^{\cF_4-\cF_2-2\cF_1}\,.
\ee
This Stokes jump is independent of the Borel transform that we chose. Taking the above jump into account in \eqref{eq:D1} and now using the $0$-Borel transform that we use throughout the paper, we find that after the Stokes transition, i.e.\ for $g$ in the left half of the complex plane, it becomes
\begin{align}
    \int_0^{-\ri \infty+\epsilon} \cB_0 [Z^{(1)}]&\left(\frac{3}{2}(\re^t-1)\right) \re^{-g t} \df t\ret
    &=   \left(1-\frac{2\re^{2\pi\ri g}}{1+\re^{4\pi \ri g}}\right)\int_0^{-\ri \infty-\epsilon} \cB_0 [Z^{(1)}]\left(\frac{3}{2}(\re^t-1)\right) \re^{-g t} \df t\,.
    \label{eq:StokesPhenB0Z1}
\end{align}
This implements the Stokes phenomenon that we use in section \ref{sec:ELORforPartitionFunction} to analytically continue the exact large order relation \eqref{eq:QuarticELOR} to \eqref{eq:QuarticELOR2} when $\frac{\pi}{2}<\gt<\frac{3\pi}{2}$.

\input{main.bbl}

\end{document}

%% file: latexdefs_all.tex
\newcommand{\gam}{\gamma}

\newcommand{\eps}{\epsilon}

\newcommand{\gt}{\theta}

\newcommand{\gl}{\lambda}
\newcommand{\gs}{\sigma}

\newcommand{\Gam}{\Gamma}

\newcommand{\gD}{\Delta}

\newcommand{\cA}{\mathcal A}
\newcommand{\cB}{\mathcal B}

\newcommand{\cF}{\mathcal F}

\newcommand{\cO}{\mathcal O}
\newcommand{\cP}{\mathcal P}

\newcommand{\cS}{\mathcal S}

\newcommand{\cZ}{\mathcal Z}

\newcommand{\bS}{\mathbb S}

\newcommand{\bZ}{\mathbb Z}

\renewcommand{\Im}{\mbox{Im}}

\renewcommand{\Re}{\mbox{Re}}

\newcommand{\be}{\begin{equation}}
\newcommand{\bea}{\begin{eqnarray}}

\newcommand{\ee}{\end{equation}}
\newcommand{\eea}{\end{eqnarray}}
\newcommand{\half}{\frac{1}{2}}

\newcommand{\ret}{\nonumber \\}

\newcommand{\mS}{\mathfrak S}

\newcommand*\df{\mathop{}\!\mathrm{d}}

\newcommand{\re}{{\rm e}}
\newcommand{\ri}{{\rm i}}

\newcommand{\fS}{{\mathsf{S}}}

\newcommand{\stir}{\genfrac\{\}{0pt}{}}

\newcommand{\nn}{\nonumber}
\def\disc{{\mathrm{disc}}}


%% file: main.bbl
\providecommand{\href}[2]{#2}\begingroup\raggedright\endgroup